\documentclass[twocolumn,aps,superscriptaddress,preprintnumbers]{revtex4}
\usepackage{graphicx}
\usepackage{amssymb}
\usepackage{epstopdf}
\usepackage{braket}
\def\KetBra#1#2{\Ket{#1}\!\!\Bra{#2}}
\usepackage{bm} 
\usepackage{longtable}
\usepackage{multirow}

\usepackage{hyperref}
\hypersetup{
    colorlinks=true,       
    linkcolor=blue,        
    citecolor=blue,        
    filecolor=blue,        
    urlcolor=blue,         
    runcolor=blue
}

\usepackage{xcolor}

\def\x{\mathbf{x}}

\def\E{\mathbf{E}}

\def\e{\mathbf{e}}
\def\bbeta{\boldsymbol{\beta}}

\def\dipole{\boldsymbol{\mu}}

\begin{document}

\title{Absorption and Photoluminescence in Organic Cavity QED}
\author{Felipe Herrera}
\email{felipe.herrera.u@usach.cl}
\affiliation{Department of Physics, Universidad de Santiago de Chile, Avenida Ecuador 3943, Santiago, Chile}
\author{Frank C. Spano}
\email{spano@temple.edu}
\affiliation{Department of Chemistry, Temple University, Philadelphia, Pennsylvania 19122, USA}

\date{\today}                                           

\begin{abstract}
Organic microcavities can be engineered to reach exotic quantum regimes of strong and ultrastrong light-matter coupling. However, the microscopic interpretation of their spectroscopic signals can be challenging due to the competition between coherent and dissipative processes involving electrons, vibrations and cavity photons. We develop here a theoretical framework based on the Holstein-Tavis-Cummings model and a Markovian treatment of dissipation to account for previously unexplained spectroscopic features of organic microcavities consistently. We identify conditions for the formation of dark vibronic polaritons, a new class of light-matter excitations that are not visible in absorption but lead to strong photoluminescence lines. We show that photon leakage from dark vibronic polaritons can be responsible for enhancing photoluminescence at the lower polariton frequency, and also explain the apparent breakdown of reciprocity between absorption and emission in the vicinity of the bare molecular transition frequency. Successful comparison with experimental data demonstrates the applicability of our theory.
\end{abstract}


\maketitle

\section{Introduction}

The experimental realization of the strong and ultrastrong coupling regimes of cavity quantum electrodynamics with organic matter  \cite{Lidzey1998,Lidzey1999,Lidzey2000,Hobson2002,Tischler2005,Holmes2007,kena-cohen2008,Kena-Cohen2010,Hutchison2013,Bellessa2014,Schartz2011,Kena-Cohen2013,Mazzeo2014,Cacciola2014,George2015-farad,Gambino2015,Long2015,Muallem2015,George2015,Shalabney2015,Saurabh2016} has stimulated interest in the development of cavity-enhanced optoelectronics \cite{Andrew2000,Hutchison2013,Feist2015,Schachenmayer2015,Orgiu2015,yuen2016}, quantum nonlinear optical devices \cite{Herrera2014,bennett2016novel,Kowalewski2016,kowalewski2016cavity},  and chemical reactors \cite{Hutchison:2012,Simpkins2015,Herrera2016}. In organic microcavities, strong light-matter coupling is possible at room temperature  for cavity quality factors as low as $Q\sim 10$ using metallic mirrors \cite{Hobson2002,George2015-farad}. These accessible conditions can facilitate experiments on the one hand, by eliminating the need to nanofabricate dielectric mirrors with high $Q$ factors, as is the case with inorganic microcavities \cite{Vahala2003}. On the other hand, organic microcavities with lossy mirrors can also obscure the interpretation of spectral measurements, as the timescales associated with coherent light-matter coupling, photon decay,  dipole decay, and vibrational motion can all be comparable.  It is thus necessary to develop a theoretical framework that can take into account these competing coherent and dissipative dynamical processes consistently, and provide a framework that can be used to interpret a wide variety of spectral measurements.

Early experiments on strong coupling with metallic microcavities by Hobson {\it et al.} \cite{Hobson2002} showed the emergence of a strong and narrow emission photoluminescence peak in a region with negligible absorption. Such results were unexpected because reciprocity dictates that a strong emitter should also be a good absorber \cite{May-Kuhn}. Since the sharp emission line was not dispersive and was close to the bare electronic transition frequency, it was attributed to  incoherent localized molecular states that could not exchange energy with the cavity field. This point of view was developed within a quasi-particle approach by Litinskaya {\it et al.} \cite{Agranovich2003,Litinskaya2004,Litinskaya2006}, who introduced the so-called incoherent exciton reservoir to partially account for the observed photoluminescence spectra \cite{Coles2011,Virgili2011}, treating the electron-vibration coupling perturbatively. However, recent experiments have identified further spectral features in the emission and absorption spectra of organic microcavities \cite{Schwartz2013,George2015-farad} that cannot be consistently explained by quasi-particle theories \cite{Agranovich2003,Litinskaya2004,Litinskaya2006,Cwik2016}. The conflicts between existing theories and experiments become more pronounced as the Rabi frequency exceeds  the intramolecular vibration frequency.

Vibrational replicas in the spectra of molecular crystals inside optical cavities were first reported by Holmes and Forrest \cite{Holmes2004}, for Rabi frequencies smaller than the intramolecular vibration frequency. Several observed features in the photoluminescence signals where captured by a quantum model developed by La Rocca {\it et al.} \cite{Mazza2009,Fontanesi2009}. Spano \cite{Spano2015} developed a numerical approach to show that allowing the admixture of multiple vibronic transitions  could lead to an effective decoupling between electron and vibrational degrees of freedom, and also reduce the effect of inhomogeneous broadening. This numerical insight was further developed and generalized in Ref. \cite{Herrera2016}, where we introduced conditions under which electronic and vibrational degrees of freedom in the lower and upper polariton states become effectively decoupled for all energy disorder configurations. The main condition for establishing this so-called {\it polaron decoupling} regime \cite{Herrera2016} is that the Rabi coupling $\sqrt{N}\Omega$ must exceed the vibronic coupling strength $\lambda^2\omega_{\rm v}$ for molecular ensembles with large $N$. Signatures of polaron decoupling in the lower polariton state could be resolved for ratios as small as $\sqrt{N}\Omega/\lambda^2\omega_{\rm v}\approx 3$, in systems with small disorder and weak vibronic coupling. One direct consequence of polaron decoupling is the reduction of the reorganization energy of excited electrons, which could be exploited to enhance electron  transfer reaction rates \cite{Herrera2016}. Subsequent variational approaches for the description of the lower polariton state  \cite{Zeb2016,Wu2016}  have confirmed the emergence of polaron decoupling in molecular ensembles. 

The theoretical tool that has improved our understanding of strongly coupled organic microcavities is the so-called Holstein-Tavis-Cummings (HTC) model, first introduced by \'{C}wik {\it et al.} \cite{Cwik2014} to study  polariton condensation with molecular ensembles. The model combines concepts from condensed matter physics \cite{Holstein:1959,Spano2010} and quantum optics \cite{Garraway2011}, by treating the quantized electronic, nuclear and photon degrees of freedom on equal footing, under the rotating-wave approximation for the cavity-matter coupling.  This fully quantized theory generalizes alternative approaches that treat the nuclear degrees of freedom classically  \cite{Galego2015,Kowalewski2016,kowalewski2016cavity}, as well as quasi-particle theories that treat the electron-vibration coupling perturbatively \cite{Agranovich2003,Litinskaya2004,Agranovich2005,Litinskaya2006}. The HTC model has a rich vibronic polariton structure in the frequency region between the conventional lower and upper polariton transmission doublet \cite{Spano2015,Herrera2016}, but in order to build a complete theoretical framework for the interpretation of organic microcavity spectroscopy, it is also necessary to treat the dissipative dynamics of organic polaritons consistently. 

In Ref. \cite{Herrera2016-PRL}, we introduce a open quantum systems approach to compute the absorption and photoluminescence spectra of organic microcavities, and successfully compare our theory with experiments. In this work, we further elaborate on the technical aspects the theory, providing additional insights into the formal structure of the HTC model, the associated vibronic polariton eigenstates, and the modelling of spectroscopic observables. We finally discuss possible improvements of the developed theory and propose experiments to test our predictions. 

\section{Holstein-Tavis-Cummings Model}

We consider an ensemble of $N$ organic emitters inside an optical cavity, which can be either individual molecules or molecular aggregates. The system can be described by the Holstein-Tavis-Cummings (HTC) Hamiltonian \cite{Cwik2014,Spano2015,Herrera2016,Zeb2016,Wu2016}, which reads
\begin{eqnarray}\label{eq:HTC}
\hat{\mathcal{H}} &=& \omega_c \,\hat a^\dagger \hat a\nonumber\\
&&+ \omega_{\rm v}\sum_{n=1}^N\hat b_n^\dagger \hat b_n+\sum_{n=1}^N\left[\omega_e+\omega_{\rm v}\lambda(\hat b_n+\hat b_n^\dagger)\right]\ket{e_n}\bra{e_n},\nonumber\\
&&+(\Omega/2)\sum_{n=1}^N(\ket{g_n}\bra{e_n}\hat a^\dagger +\ket{e_n}\bra{g_n}\hat a),
\end{eqnarray}
where $\omega_e =\omega_{00}+ \omega_{\rm v}\lambda^2$ is the vertical Frank-Condon transition frequency, with $\omega_{00}$ being the frequency of the zero-phonon (0-0) vibronic transition, $\omega_{\rm v}$ is the intramolecular vibrational frequency and $\lambda^2$ is the Huang-Rhys factor \cite{Spano2010}, which quantifies the strength of vibronic coupling. The operator $\hat b_n$ annihilates one quantum of vibration on the $n$-th chomophore. The operator $\hat a$  annihilates a cavity photon at frequency $\omega_c$, and $\Omega$ is the single-particle vacuum Rabi frequency, which can be on the order of $100$ meV in plasmonic nanocavities \cite{Chikkaraddy:2016aa}. In microcavities, collective couplings $\sqrt{N}\Omega\sim 0.7 -1.0$ eV have been reported \cite{Schwartz2011,Kena-Cohen2013}.  

In microcavities, the photon frequency $\omega_c$ depends on the in-plane wave vector $k_\parallel$ of the cavity mode. The HTC model in Eq. (\ref{eq:HTC}) implicitly assumes that all molecular transitions couple equally strong to photons having a specific value of $k_\parallel$. This can only be justified for lossy microcavities near $k_\parallel =0$ (normal incidence), where the photon dispersion cannot be resolved within the cavity linewidth. Despite this simplification, we will consider the polariton structure of the HTC model for higher in-plane momenta $k_\parallel > 0$ in later sections.  Throughout this work,  we assume that the cavity detuning, defined as $\Delta \equiv \omega_{00}-\omega_{c}$, vanishes exactly at normal incidence. 

We ignore energetic disorder in Eq. (\ref{eq:HTC}), since for parameters consistent with experimental data  \cite{Hobson2002,Coles2011,Virgili2011,Schwartz2013}, we find that the inhomogeneous broadening associated with energy disorder does not contribute significantly to the overall absorption and emission lineshapes \cite{Herrera2016-PRL}. Within the HTC model,  all molecules in the ensemble couple equally strong to the electric field of the near-resonant cavity mode $\E_c$, which ignores the possibility of molecules being oriented in the sample such that the product $\dipole \cdot \E_c$ is negligibly small. We expect such uncoupled molecules to behave as free-space emitters, possibly exhibiting a cavity-enhanced emission rate typical of the weak cavity-matter coupling regime \cite{Barnes1998}. The theory we develop in the following sections applies only to those emitters in the sample for which the HTC model provides an accurate description of their photophysics. 

\subsection{Symmetry of the resonant HTC model}
\label{sec:symmetry} 

Before proceeding with the detailed analysis of the eigenstates of the HTC model, we introduce a unique symmetry transformation $\hat S$ that is shown to commute with the HTC Hamiltonian $\hat {\mathcal{H}}$ in Eq. (\ref{eq:HTC}).  In this Section we define the transformation $\hat S$ and prove some of its basic properties for the single-particle and many-particle cases.

\begin{figure}[t]
\begin{center}
\includegraphics[width=0.47\textwidth]{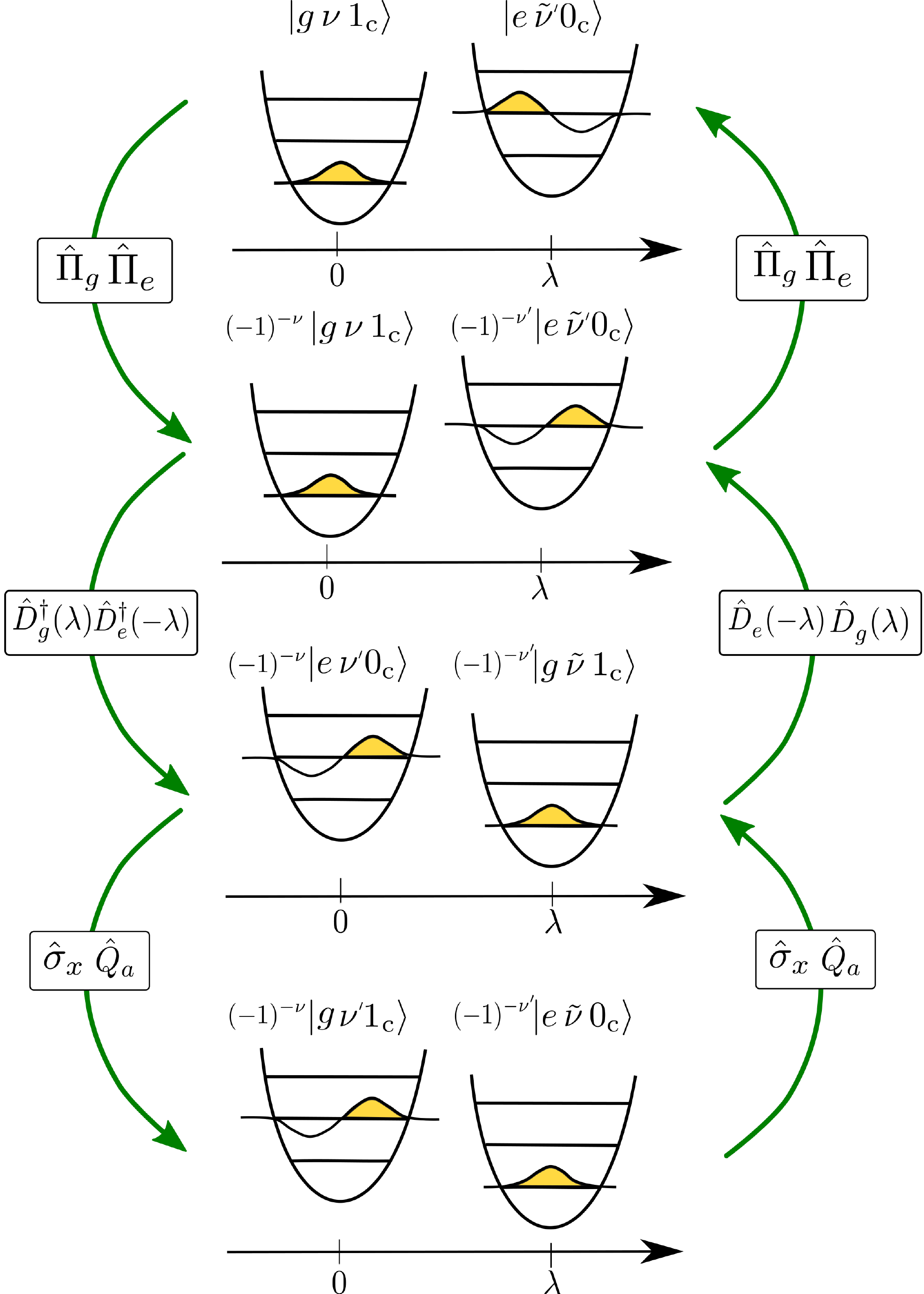}
\end{center}
\caption{Step-wise illustration of the symmetry transformation $\hat S$ on an arbitrary vibronic configuration involving the ground and excited electronic potentials. The downward arrows represent the action of $\hat S$ and the upward arrows its inverse. The horizontal axis represent the dimensionless vibrational coordinate.}
\label{fig:symmetry}
\end{figure}

For a single emitter ($N=1$), we can rewrite the light-matter coupling term of the HTC Hamiltonian $\hat{\mathcal{H}}$ by expanding the dipole transition and cavity field operators in the basis spanned by the states $\ket{g\,\nu,1_c}$ and $\ket{e\,\tilde \nu\,0_c}$, which represent, respectively, a molecule in its ground electronic state $\ket{g}$ with vibrational eigenstate $\ket{\nu}$ and one cavity photon, and a molecule in the excited electronic state $\ket{e}$, with vibrational eigenstate $\ket{\tilde \nu}$ in the cavity vacuum. The tilde overstrike indicates that the vibrational eigenstate has $\tilde\nu$ vibrational quanta in the excited state harmonic nuclear potential, whose minimum is shifted relative to the ground state potential minimum by a quantity $\lambda$ along a dimensionless vibrational coordinate \cite{May-Kuhn,Spano2010}. The resulting light-matter interaction Hamiltonian reads
\begin{equation}\label{eq:HLM}
\hat H_{\rm LM} = \frac{1}{2}\sum_{\nu\nu'}\Omega_{\tilde \nu' \nu}\KetBra{e\tilde \nu' 0_c}{g\nu 1_c}+\Omega_{\nu\tilde\nu'}\KetBra{g\nu 1_c}{e\tilde \nu' 0_c},
\end{equation}
where $\Omega_{\tilde\nu' \nu} = \langle \tilde \nu'|\nu \rangle \Omega$ is a vibronic Rabi frequency, weighted by the vibrational overlap factor $\langle \tilde \nu'|\nu \rangle $. We introduce the composite symmetry transformation $\hat S$, defined as
\begin{equation}\label{eq:S}
\hat S = \hat \sigma_x\hat Q_a\hat D^\dagger_g(\lambda)\hat D^\dagger_e(-\lambda)\hat \Pi_g\hat \Pi_e,
\end{equation}
where $\hat \Pi_g$ and $\hat \Pi_e$ are the usual parity and displaced parity operators 
\begin{equation}\label{eq:Pi}
\hat \Pi_g  ={\rm e}^{i\pi b^\dagger  b}\;\;,\;\;\hat \Pi_e  ={\rm e}^{i\pi \tilde b^\dagger \tilde b},
\end{equation}
acting on vibrational eigenstates of the ground and excited state potentials as $\hat \Pi_g\ket{\nu} = (-1)^{\nu}\ket{\nu}$ and  $\hat \Pi_e\ket{\tilde \nu'} = (-1)^{\nu'}\ket{\tilde \nu'}$. The displacement operators have the usual definition (${\rm Im}[\lambda] = 0$)
\begin{equation}\label{eq:D}
\hat D (\lambda) = {\rm e}^{\lambda (b^\dagger -b )}, 
\end{equation}
such that displaced harmonic oscillator eigenstates are defined as $\ket{\tilde \nu} = \hat D^\dagger (\lambda)\ket{\nu}$ \cite{May-Kuhn}.
The definition of $\hat S$ also includes the Pauli matrix $\sigma_x$ in the $\{\ket{e},\ket{g}\}$ basis, i.e., $\hat \sigma_x = \ket{g}\bra{e}+\ket{e}\bra{g}$ as well as the one-photon cavity quadrature operator $\hat Q_a = \ket{1_c}\bra{0_c}+\ket{0_c}\bra{1_c}$, in the truncated Fock space. By construction, the symmetry transformation $\hat S$ is unitary, i.e., $\hat S^\dagger \hat S = 1$, as illustrated in Fig. \ref{fig:symmetry}.

We prove in Appendix \ref{app:symmetry} that the transformation $\hat S$ commutes with the light-matter term $\hat H_{\rm LM}$.  Therefore, an eigenstate $\ket{\epsilon_j}$ of the light-matter term $\hat H_{LM}$ is a simultaneous eigenstate of $\hat S$, satisfying the eigenvalue equation $
\hat S\ket{\epsilon_j} = s_j\ket{\epsilon_j}$, with $s_j^2=1$. We can thus classify the eigenstates of $\hat H_{\rm LM}$ according to the symmetry eigenvalue $s_j$. However, the total system Hamiltonian $\hat{\mathcal{H}}$ also includes diagonal terms that describe the bare cavity photon and molecular degrees of freedom, and their commutation properties with the symmetry transformation $\hat S$ also need to be established. These diagonal terms can be written in the basis described above as  
\begin{eqnarray}\label{eq:H diagonal}
\hat{\mathcal{ H}}-\hat H_{\rm LM} &=& \sum_\nu (\omega_c+ n_\nu\omega_{v})\, \ket{g\,\nu\,1_c}\bra{g\,\nu\,1_c}\nonumber\\
&&+\sum_{\nu} (\omega_e+n_{\tilde \nu }\omega_v)\ket{e\,\tilde \nu\,0_c}\bra{e\,\tilde \nu\,0_c}
\end{eqnarray}
where $n_\nu$ and $n_{\tilde \nu'}$ are the number of vibrational quanta (phonons) in the ground and electronic harmonic nuclear potentials, assumed to have equal vibrational frequencies.  

Equation (\ref{eq:H diagonal}) shows that under conditions of exact cavity-matter resonance, i.e., $\omega_e = \omega_c$, we also have \[[ (\hat{\mathcal{ H}}-\hat H_{\rm LM}), \hat S] = 0 = [\hat{\mathcal{ H}}, \hat S]. \] In other words, under exact resonance the eigenstates of the HTC model for $N=1$ are also eigenstates of the symmetry transformation $\hat S$, and the HTC Hamiltonian matrix can be factorized into block-diagonal sub-matrices that conserve the symmetry eigenvalue $s$.  

For many-particle systems $(N\geq 2)$, the internal degrees of freedom of the $n$-th particle in the ensemble can be still be associated with a symmetry transformation $\hat S_n$ of the form in Eq. (\ref{eq:S}). Many-particle eigenstates of the HTC model $\ket{\epsilon_j}$ can thus be classified according to the composite transformation $\hat S_\alpha$ that takes into account the symmetry of the many-body electronic state under permutations of the molecules in the ensemble. One possible choice relevant for the one-excitation manifold is 
\begin{equation}\label{eq:S alpha}
\hat S_\alpha = \sum_{n=1}^N c_{\alpha n} \hat S_{n},
\end{equation}
where the set of orthonormal coefficients $c_{\alpha n}$ encode the symmetry under particle permutations, characterized the  quantum number $\alpha$.  

\subsection{Symmetry Classification of Diabatic Vibronic Polaritons}

For individual particles ($N=1$), it proves convenient to introduce diabatic vibronic polaritons of the form 
\begin{equation}\label{eq:diabatic}
\ket{\nu_\pm} = \frac{1}{\sqrt{2}}\left(\ket{e\,\tilde \nu \, 0_c}\pm \ket{g\,\nu\,1_c}\right), 
\end{equation}
which become accurate polariton eigenstates of the system Hamiltonian $\hat{\mathcal{H}}$ for $\sqrt{N}\Omega/\omega_{\rm v}\ll 1$ \cite{Michetti2008,Mazza2009}.  
From the definition of $\hat S$ in Eq. (\ref{eq:S}), the following relations hold 
\begin{eqnarray}\label{eq:polariton relations}
\hat S\ket{\nu_+} &=& (-1)^\nu \ket{\nu_+}\\
\hat S\ket{\nu_-} &=& (-1)^{\nu+1} \ket{\nu_-}, 
\end{eqnarray}
which are eigenvalue relations with eigenvalues $s_j =\pm 1$, depending on the vibrational configuration. We refer to those diabatic states with eigenvalue $s = 1$ as {\it even} vibronic polaritons and those with $s=-1$ as {\it odd} vibronic polaritons.

As an example, let us consider the system Hamiltonian $\hat{\mathcal{H}}$  for $N=1$, in a truncated subspace spanned by the diabatic polariton states $\{\ket{0_\pm}, \ket{1_\pm}, \ket{2_\pm}\}$. This basis set splits according to the symmetry eigenvalue $s$ into an even subspace $\mathcal{S}_+ = \{\ket{0_+},\ket{1_-},\ket{2_+}\}$  and an odd subspace $\mathcal{S}_- = \{\ket{0_-},\ket{1_+},\ket{2_-}\}$. 
The single-molecule Hamiltonian matrix under exact resonance ($\omega_c =\omega_{00}$) can thus be written in the block-diagonal form
\begin{eqnarray}\label{eq:blocks}
\lefteqn{\hat{\mathcal{H}}/\omega_{\rm v} = }\nonumber\\
&&\left(\begin{array}{cccccc}
\bar\Omega_{0\tilde 0}&\bar\Omega_{0\tilde 1}&\bar\Omega_{0\tilde 2}&0&0&0\\
\bar\Omega_{0\tilde 1}&1-\bar\Omega_{1\tilde 1}&\bar\Omega_{1\tilde 2}&0&0&0\\
\bar\Omega_{0\tilde 2}&\bar\Omega_{1\tilde 2}&2+\bar\Omega_{2\tilde 2}&0&0&0\\
0&0&0&-\bar\Omega_{0\tilde 0}&\bar\Omega_{0\tilde 1}&\bar\Omega_{0\tilde 2}\\
0&0&0&\bar\Omega_{0\tilde 1}&1+\bar\Omega_{1\tilde 1}&\bar\Omega_{1\tilde 2}\\
0&0&0&\bar\Omega_{0\tilde 2}&\bar\Omega_{1\tilde 2}&2-\bar\Omega_{2\tilde 2}
\end{array}\right),\nonumber\\
\end{eqnarray}
where $\bar\Omega_{\nu\tilde\nu}=(\Omega/2\omega_{\rm v})\,\langle \nu|\tilde\nu\rangle$ are vibronic Rabi frequencies in units of the vibrational frequency $\omega_{\rm v}$. We have set the zero of energy to be at the cavity frequency $\omega_c$, which corresponds to a rotating frame transformation \cite{Barnett-Radmore}.  

The even parity sub-block in Eq. (\ref{eq:blocks}) (on the upper left) includes the conventional diabatic upper polariton. As we discuss below, this sub-block supports a zero-energy eigenvalue. The odd parity sub-block (lower right), involves the conventional lower polariton. The off-diagonal couplings within a given sub-block admix diabatic polaritons associated with different molecular vibronic transitions, such that the number of vibrational quanta in the ground and excited state harmonic nuclear potentials is not conserved. This admixture becomes significant for Rabi couplings  $\Omega/\omega_{\rm v}\approx 2$.

\section{Dark Vibronic Polaritons}
\label{sec:DVP}

We introduce the concept of a dark vibronic polariton \cite{Herrera2016-PRL} to describe eigenstates of the Holstein-Tavis-Cummings model $\ket{\epsilon_j}$ for which we have
\begin{equation}\label{eq:mu j}
\mu_{Gj}=\langle G|\hat \mu|\epsilon_j \rangle = 0,
\end{equation}
with $\ket{G}\equiv \ket{g_1\,0_1,g_2\,0_2,\ldots,g_N\,0_N}\ket{0_c}$ being the absolute ground state of the organic cavity, having no electronic, vibrational or cavity excitations, for a dipole operator written as $\hat \mu = \mu(\hat \mu^{(+)}+\hat \mu^{(-)})$, where 
\begin{equation}\label{eq:mu}
\hat \mu^{(+)} = \sum_{n=1}^N\,\ket{g_n}\bra{e_n}, 
\end{equation}
with $\hat\mu^{(-)}=[\hat \mu^{(+)}]^\dagger$. The molecular transition dipole moment $\mu$  is assumed identical for each emitter.  
 Absorption of light from bound dielectric or surface modes of the microcavity \cite{Barnes1998,Matterson2001} by the $j$-th polariton eigenstate is proportional to $|\mu_{Gj}|^2$, just as in free space spectroscopy \cite{Spano2010}.  Dark vibronic polaritons are therefore invisible in bound mode absorption at their eigenfrequencies $\omega_j$, which does not prevent them from emitting light in photoluminescence, as we discuss in detail below.   

We identify two types of dark vibronic polaritons: $X$ and $Y$ types \cite{Herrera2016-PRL}. The main distinguishing feature between these two types of polariton eigenstates is the degeneracy of their eigenfrequencies: while $X$-type vibronic polaritons are non-degenerate, $Y$-type states are $(N-1)$-fold degenerate. We show below that although $X$-type dark vibronic polaritons can be partially visible in conventional microcavity absorption measurements, dark vibronic polaritons of the $Y$ type remain invisible for any type of absorption measurement, as long as $k_{\rm b}T/\omega_{\rm v}\ll 1$, which is typical at room temperature.

\subsection{Single Particle States}
\label{sec:SP}

\begin{figure}[t]
\includegraphics[width=0.4\textwidth]{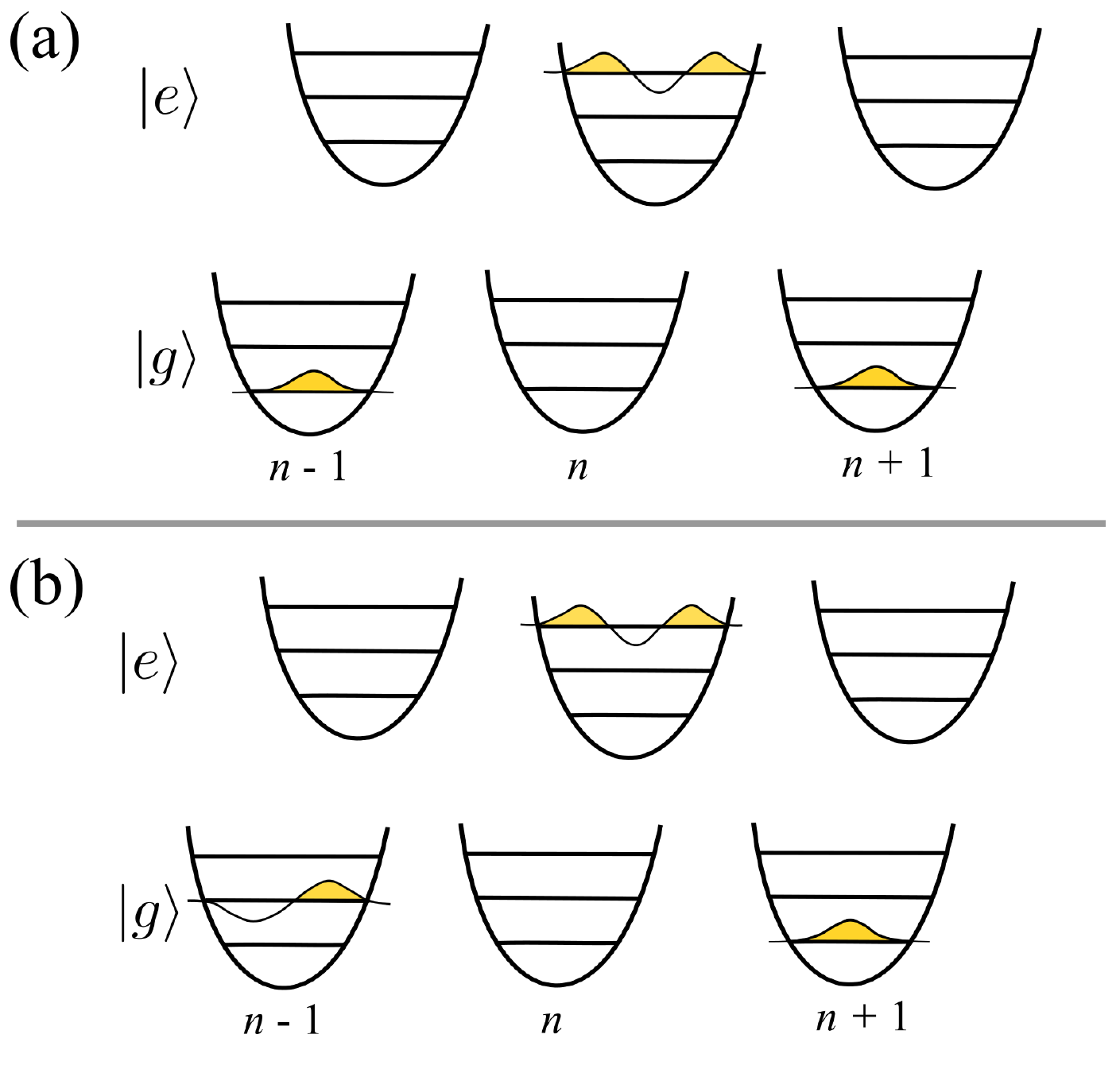}
\caption{(a) Displaced-potential representation of a single-particle material state, having a vibronic excitation $\ket{e_n\tilde \nu_n =2}$ on the $n$-th molecule of the ensemble.  (b)  Two-particle material state with a vibronic excitation ($\tilde \nu =2$) on molecule $n$  and a vibrational excitation ($\nu=1$) on molecule $n-1$. }
\label{fig:SP-TP}
\end{figure}

Before discussing the structure of dark vibronic polaritons of the $X$ and $Y$ type in more detail, we need to define diabatic single-particle and two-particle material and polariton states. These are illustrated in Fig. \ref{fig:SP-TP}. Single particle {\it material} states represent a molecular ensemble hosting a single vibronic {\it or} vibrational excitation, or in the language of ordered molecular aggregates, a single exciton or  phonon. We represent single particle vibronic material states as
\begin{equation}\label{eq:SP material}
\ket{\alpha,\tilde\nu,0_c} = \sum_{n=1}^N c_{\alpha n}\ket{g_10_1,\ldots,e_n\,\tilde \nu_n\ldots g_N0_N,0_c},
\end{equation}
where $\alpha$ is a quantum number associated with permutation symmetry of the electronic degrees of freedom, such that the expansion coefficients satisfy $|c_{\alpha n}|^2=1/N$, i.e., all molecules are equally likely to host the vibronic excitation $\tilde \nu$ in state $\ket{e}$. There are thus $N$ possible values of $\alpha$. In the language of molecular aggregates with translational invariance \cite{Spano2010}, the quantum number $\alpha$ coincides with the exciton quasi-momentum $k$, which can take $N$ possible values given by multiple integers of $2\pi/Na_L$, where $a_L$ is the lattice constant. For organic cavities however, there is no such translational symmetry as each emitter is randomly located in the sample. Permutation symmetry thus emerges from the fact that within the Holstein-Tavis-Cummings model, all molecules in the ensemble are equally coupled to the confined electric field of the microcavity. In quantum optics, this permutation symmetry is also known to emerge from the Tavis-Cummings model with homogeneous couplings \cite{Garraway2011}, when represented in the collective spin angular momentum basis \cite{Carmichael-book1}. 

Following Ref. \cite{Herrera2016}, we distinguish states whose electronic degrees of freedom are totally symmetric with respect to particle permutation by the quantum number $\alpha=\alpha_0$, such that $c_{\alpha_0 n} = 1/\sqrt{N}$ for all $n$. For molecular aggregates, this would correspond to a $k=0$ exciton, which is superradiant for J-aggregates \cite{Spano1989a}. In the HTC model, only single particle states with  quantum number $\alpha_0$ can directly exchange energy with the cavity field, under the assumption of homogeneous Rabi coupling. There are thus $N-1$ single-particle states with $\alpha\neq \alpha_0$ that are not totally symmetric with respect to permutations and thus do not couple directly to the cavity field. These states are the cavity analogues of the $k\neq 0$ excitons in molecular aggregates, which do not couple to light directly, but can be involved in non-radiative processes  \cite{Spano2010}. 

The Holstein-Tavis-Cummings model in Eq. (\ref{eq:HTC}) can resonantly couple the totally symmetric single-particle material states $\ket{\alpha_0,\tilde \nu,0_c}$ with a vibrationless single photon state to give a diabatic vibronic {\it polariton} state of the form 
\begin{equation}\label{eq:SP polariton}
\ket{\alpha_0, \tilde \nu,\pm} = \frac{1}{\sqrt{2}}\left(\ket{\alpha_0,\tilde \nu,0_c}\pm\ket{g_10_1,g_20_2\ldots,g_N0_N,1_c}\right), 
\end{equation}
with diabatic energies
\begin{equation}\label{eq:SP doublet}
E_{\tilde \nu}^\pm  = \tilde \nu \omega_{\rm v}\pm \sqrt{N}\Omega|\langle 0|\tilde \nu\rangle |/2.
\end{equation}
We recover the conventional lower and upper polariton states for $\tilde \nu =0$ \cite{Spano2015,Herrera2016},  associated with the polariton splitting $(E_{\tilde 0}^{+}-E_{\tilde 0}^{-}) =  \sqrt{N}\Omega|\langle 0 |\tilde 0\rangle |$.

\subsection{Two-Particle States}
\label{sec:TP}

Two particle states represent a configuration where one molecule in the ensemble hosts a vibronic excitation in state $\ket{e}$, while a different molecule hosts a vibrational excitation in state $\ket{g}$. Two-particle states are  important to describe the optical spectra of molecular aggregates in free space \cite{Spano2010} and in microcavities \cite{Spano2015}. Three-particle and multi-particle states can be defined in a similar way, but for the Rabi coupling strengths of interest in this work  $1\leq  \sqrt{N}\Omega/\omega_{\rm v}\leq 3$, their contribution to the absorption and emission spectra in the frequency region within the conventional lower and upper polariton doublet is negligible. For larger Rabi couplings, possibly in the ultrastrong coupling regime \cite{Cacciola2014,Gambino2015,Kena-Cohen2013,Mazzeo2014,Schartz2011}, such mutli-particle excitations should not be ignored \cite{Herrera2016}. 

In general, we can represent two-particle vibronic-vibrational material states as
\begin{eqnarray}\label{eq:TP material}
\lefteqn{\ket{\alpha\beta,\tilde \nu' \nu,0_c} =}\\
&&  \sum_{n\neq m}\sum_mA_{\alpha n }^{\beta m}\ket{g_10_1,\ldots,e_n\tilde \nu'_n,\ldots, g_m\nu_m,\ldots,g_N0_N,0_c}\nonumber,
\end{eqnarray}
where the amplitude $A_{\alpha n }^{\beta m}$ is defined by the permutation symmetry quantum number of the electronic degrees of freedom $\alpha$, and also the permutation quantum number of the vibrational configuration $\beta$. The definition in Eq. (\ref{eq:TP material}) involves $N(N-1)$ possible vibronic-vibrational configurations. 

Let us write the light-matter term of the HTC model $\hat H_{\rm LM}$ in terms of the totally-symmetric electronic state  $\ket{\alpha_0} = \sum_{l=1}^N\ket{e_l}/\sqrt{N}$ as \cite{Herrera2016} 
\begin{equation}\label{eq:HLM alphazero}
\hat H_{\rm LM} = \frac{\sqrt{N}\Omega}{2}\left(\ket{\alpha_0}\bra{g_1,g_2\ldots,g_N}\hat a +{\rm H.c.}\right), 
\end{equation}
where we have simply changed to a collective electronic basis in Eq. (\ref{eq:HTC}). We can use this form of $\hat H_{\rm LM}$ to understand what photonic states couple to the two-particle material states defined in Eq. (\ref{eq:TP material}). For example, we have
\begin{equation}
\bra{g_10_1,\ldots,g_N0_N,1_c}\hat H_{\rm LM}\ket{\alpha\beta,\tilde \nu'\nu,0_c} = 0, 
\end{equation}
which by the arguments in Appendix \ref{app:muX} implies that the state $\ket{\alpha\beta,\tilde \nu'\nu,0_c}$ has no transition dipole moment to the absolute ground state of the cavity. In general, two-particle material states can couple to single photon states having $\nu\geq 1$ vibrational excitations, given by
\begin{equation}\label{eq:SP nu}
\ket{ \beta,\nu,1_c} =\sum_{n=1}^Nc_{\beta n}\ket{g_10_1\ldots,g_n\nu_n,\ldots,g_N0_N,1_c},
\end{equation}
representing a single cavity photon state dressed by a collective vibrational excitation with permutation quantum number $\beta$. The associated light-matter coupling element is given by
\begin{equation}\label{eq:SPTP coupling}
\bra{\beta',\nu,1_c}\hat H_{\rm LM} \ket{\alpha\beta,\tilde \nu'\nu,0_c} = \frac{\Omega}{2}\langle 0|\tilde \nu'\rangle \sum_{m}\sum_{l\neq m}c^*_{\beta'm} A_{\alpha l}^{\beta m},
\end{equation}
where the double summation contains $N(N-1)$ terms. This expression can be considered as a type of selection rule for the light-matter coupling between two-particle material states and vibrationally-dressed photon states. For simplicity, we use the ansatz 
\begin{equation}\label{eq:TP ansatz}
A_{\alpha n}^{\beta m} = \frac{c_{\beta m}}{\sqrt{N-1}},
\end{equation}
for all $n$., i.e., the vibronic configurations are totally symmetric under particle permutations. This form further simplifies Eq. (\ref{eq:SPTP coupling}) to read
\begin{equation}\label{eq:ansatz coupling}
\bra{\beta',\nu,1_c}\hat H_{\rm LM} \ket{\alpha_0\beta,\tilde \nu'\nu,0_c} = \sqrt{N-1}\left(\frac{\Omega}{2}\right)\langle 0|\tilde \nu'\rangle \delta_{\beta \beta'}.
\end{equation}
The kronecker delta factor $\delta_{\beta\beta'}$ is the  cavity analogue of phonon momentum conservation in systems with translational invariance.  

The coupling in Eq. (\ref{eq:ansatz coupling}) between two-particle {\it material} states $\ket{\alpha\beta,\tilde \nu'\nu,0_c}$ and vibrationally-excited single photon states of the cavity, give rise to diabatic two-particle {\it polariton} states \cite{Spano2015} of the form
\begin{equation}\label{eq:TP polariton}
\ket{P_{ \nu \tilde\nu'}^\pm,\beta} = \frac{1}{\sqrt{2}}\left(\ket{\alpha_0\beta,\tilde\nu'\nu,0_c}\pm \ket{\beta,\nu,1_c}\right),
\end{equation}
where the first term in the superposition is given by Eq. (\ref{eq:TP material}) for the separable ansatz in Eq. (\ref{eq:TP ansatz}).  On resonance ($\omega_c = \omega_{00}$), these two-particle polaritons have the diabatic energies
\begin{equation}\label{eq:Enu}
E_{\nu\tilde \nu'}^{\pm} = (\tilde \nu'+\nu)\omega_{\rm v}\pm \sqrt{N-1}\,\Omega\,|\langle 0|\tilde \nu' \rangle|/2,
\end{equation} 
which split into a lower (minus sign)  and upper upper (plus sign) diabatic polariton manifold, each being $N$-fold degenerate according to the $N$ energetically equivalent choices of the permutation quantum number $\beta$. The diabatic two-particle polaritons in Eq. (\ref{eq:TP polariton}) can become accurate eigenstates of the HTC Hamiltonian for $\sqrt{N}\Omega/\omega_{\rm v}\ll 1$.

For the symmetric quantum number $\beta =\alpha_0$, the two-particle diabatic vibronic polariton $\ket{P_{ \nu \tilde\nu'}^\pm,\alpha_0}$ [Eq. (\ref{eq:TP polariton}) can couple to conventional single-particle polaritons $\ket{\alpha_0,\tilde \nu',\pm}$ [Eq. (\ref{eq:SP polariton})] by the light-matter term according to the matrix element
\begin{equation}\label{eq:pm couplings}
\bra{\alpha_0,\tilde \nu',\pm}\hat H_{\rm LM}\ket{P_{ \nu \tilde\nu'}^\pm,\alpha_0} = \frac{\Omega}{4}\langle \tilde \nu'|\nu \rangle,
\end{equation}
independent of the number of molecules $N$.

We conclude this section by noting that the Holstein-Tavis-Cummings model also allows the coupling between single-particle material states $\ket{\alpha\neq \alpha_0,\tilde \nu,0_c}$ [Eq. (\ref{eq:SP material})] that are not totally-symmetric with respect to permutation, with vibrationally-excited photon states $\ket{\beta, \nu,1_c}$. Such coupling preserves the permutation symmetry quantum number, i.e., 
\begin{equation}\label{eq:alpha=beta}
\alpha = \beta\neq \alpha_0.
\end{equation}
resulting in the non-zero HTC matrix element 
\begin{equation}\label{eq:SP-TP coupling}
\bra{\alpha,\tilde \nu',0_c}\hat H_{\rm LM}\ket{P_{ \nu \tilde\nu'}^\pm,\beta} = \delta_{\beta\alpha} \frac{\langle \nu|\tilde \nu'\rangle}{\sqrt{2}}\left(\frac{\Omega}{2}\right),
\end{equation} 
also independent of $N$. This expression is valid for the $(N-1)$ possible values of $\alpha\neq \alpha_0$ (non-symmetric superpositions). As a consequence of this coupling, the otherwise dark collective electronic excitations $\ket{\alpha\neq \alpha_0,\tilde \nu,0_c}$ can admix with two-particle polariton states $\ket{P_{\nu\tilde \nu'}^\pm, \alpha}$ given by Eq. (\ref{eq:TP polariton}), to borrow a photonic component with $\nu\geq 1$ vibrational quanta. As we discuss below, such admixture can be significant in the Rabi coupling regime $\sqrt{N}\Omega/\omega_{\rm v}\approx 2$, where the states involved in the mixing become quasi-degenerate. 
 
\subsection{Dark Vibronic Polaritons of the $X$ Type}
\label{sec:X type}

\begin{figure}[t]
\includegraphics[width=0.43\textwidth]{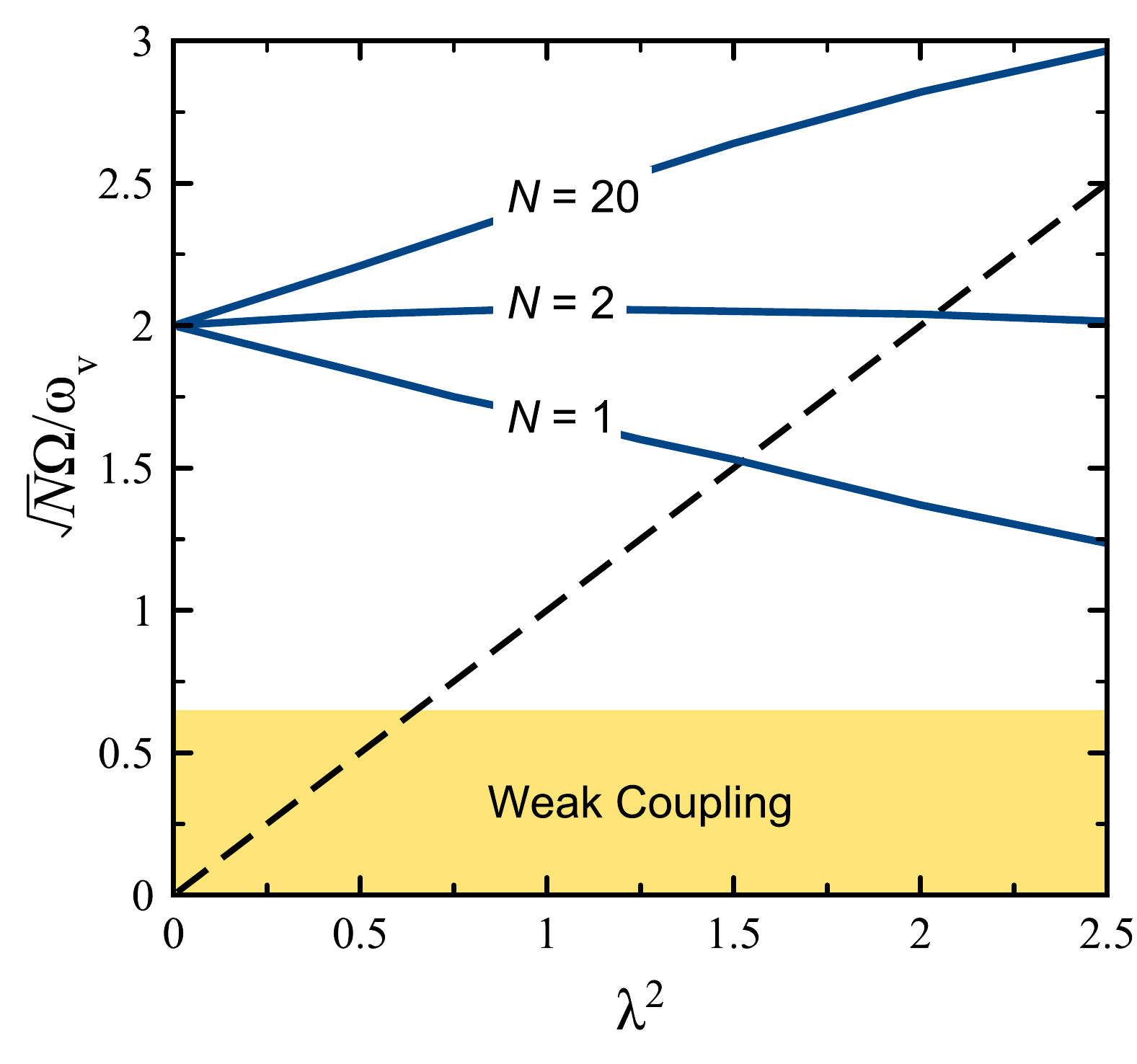}
\caption{HTC model parameters. Solid lines are the parameters for which we can define the zero energy dark vibronic polariton $\ket{X}$, for different molecule numbers $N$, where $N=20$ is representative of the thermodynamic limit \cite{Zeb2016}.  The dashed line shows the threshold for  polaron decoupling of the lower polariton state, defined as $\sqrt{N}\Omega/\lambda^2\omega_{\rm v} = 1$ for $N\gg 1$.  $\lambda^2$ is the Huang-Rhys factor and $\omega_{\rm v}$ the intramolecular frequency. The shaded region represents the weak coupling regime $\sqrt{N}\Omega\leq \kappa$, where $\kappa\sim 0.6\,\omega_{\rm v}$ is a typical photon decay rate for a microcavity with quality factor $Q\sim 10$ \cite{Hobson2002}. The cavity is assumed resonant with the zero-phonon electronic transition at normal incidence.}
\label{fig:HTC params}
\end{figure}

We are interested in the eigenstates of the HTC Hamiltonian $\hat{\mathcal{H}}$  [Eq. (\ref{eq:HTC})] which have energies close to the bare molecular transition frequency, given the unexpectedly strong cavity emission and weak absorption observed in this frequency region \cite{Hobson2002,Coles2011,Virgili2011,Schwartz2013}. In the rotating frame of the resonant cavity mode, a polariton state at the bare molecular frequency satisfies 
\begin{equation}\label{eq:HX0}
\hat{\mathcal{H}}\ket{X}=0.
\end{equation}
We provide a simple proof in Appendix \ref{app:muX} for the fact that Eq. (\ref{eq:HX0}) implies also a vanishing transition dipole moment $\mu_{GX}$, as defined in Eq. (\ref{eq:mu j}). State $\ket{X}$ is thus a dark vibronic polariton, which we call of the $X$ type. We show in Fig. \ref{fig:HTC params} the numerically estimated values of the HTC Hamiltonian parameters $\sqrt{N}\Omega/\omega_{\rm v}$ and $\lambda^2$ for which one of the eigenstates of $\hat{\mathcal{H}}$ satisfies Eq. (\ref{eq:HX0}), without degeneracies. For completeness, we also shown in Fig. \ref{fig:HTC params} what we define as the polaron-decoupling threshold of the lower polariton state in molecular ensembles \cite{Herrera2016},  defined by the equality 
\[\sqrt{N}\Omega/\lambda^2\omega_{\rm v}=1.\] 
For Rabi couplings a few times higher than this threshold, the electronic and vibrational degrees of freedom  of the lower polariton state become separable, with deviations from complete separability that scale as $\lambda^2/4N$ \cite{Herrera2016}.

We can understand the emergence of the zero-energy dark vibronic polariton $\ket{X}$ as a destructive interference effect. For $N=1$, it is not difficult to show that only the even-symmetry sub-block in Eq. (\ref{eq:blocks}) can have a zero-energy eigenvalue. Therefore, we can expand the $\ket{X}$ state in a diabatic polariton basis from Eq. (\ref{eq:diabatic}) to read
\begin{equation}\label{eq:X state}
\ket{X}\approx c_0\ket{0_+}-c_1\ket{1_-}-c_2\ket{2_+}
\end{equation}
 where $c_{\nu}> c_{\nu+1}>0$, and $\lambda>0$. We have assumed the same even-symmetry truncated basis as in Eq. (\ref{eq:blocks}).  The transition dipole moment $\mu_{GX}$ is thus given by
\begin{equation}\label{eq:muX}
\mu_X \approx \mu \left(c_0\langle 0|\tilde 0\rangle -c_1\langle 0 |\tilde 1\rangle-c_2\langle 0 |\tilde 2\rangle\right ),
\end{equation}
i.e., the destructive interference between the vibrational overlap factors associated with the bare 0-0, 0-1 and 0-2 vibronic transitions leads to the vanishing dipole moment for state $\ket{X}$. The destructive interference is more efficient when higher energy polariton states are considered in the basis set. 

Figure \ref{fig:HTC params} shows that for any $N$ it is always possible to find values of the HTC model parameters for which  Eq. (\ref{eq:HX0}) holds, which generalizes the definition of the $\ket{X}$ state in Eq. (\ref{eq:X state}) to the many-particle case. We note that for $N\geq 2$, the zero-energy $\ket{X}$ state also acquires a two-particle polariton character, due to the coupling between the conventional upper polariton state $\ket{\alpha_0,\tilde 0,+ }$ with the symmetric two-particle polariton state $\ket{P_{1\tilde 0}^-,\alpha_0}$. This type of coupling, in addition to the admixture of single-particle polariton states as in Eq. (\ref{eq:X state}), gives rise to a set of {\it non-degenerate} dark vibronic polaritons that we call of the $X$ type in the frequency region defined by the conventional lower and upper polariton doublet.

\subsection{Dark Vibronic Polaritons of the $Y$ Type}
\label{sec:Y type}

Dark vibronic polaritons of the $Y$ type result from the admixture of diabatic two-particle polariton states $\ket{P_{\nu\tilde \nu'}^\pm,\beta}$, [Eq. (\ref{eq:TP polariton})] with non-symmetric (or dark) material vibronic excitations $\ket{\beta,\tilde \nu',0_c}$, which is allowed by the light-matter term $\hat H_{\rm LM}$ in the HTC model [Eq. (\ref{eq:SP-TP coupling})]. Since there are $N-1$ possible values of the non-symmetric permutation quantum number $\beta\neq \alpha_0$, the resulting HTC polariton eigenstates split into two $(N-1)$-fold degenerate manifolds, involving  upper (plus sign) and lower (minus) two-particle polariton states. 

For Rabi frequencies $\sqrt{N}\Omega/\omega_{\rm v}\approx 2$, where we expect to form dark vibronic polaritons (see Fig. \ref{fig:HTC params}),  there are $Y$-type polariton eigenstates in the spectral region near the bare electronic frequency $(\omega_j\approx 0)$ that can be approximately written as 
 \begin{equation}\label{eq:Y type}
 \ket{Y_j} \approx a_{j}\ket{\beta,\tilde 0, 0_c}+b_{j}\ket{P_{1\tilde 0}^-,\beta},
 \end{equation} 
with real coefficients $a_j\sim b_j$. States of this form become important in the description of the photoluminescence spectra.

\section{Dissipative Polariton Dynamics }

In order to model the dissipative dynamics of organic polaritons in low-$Q$ microcavities, we ignore non-radiative intramolecular relaxation \cite{May-Kuhn} and assume the dissipative dynamics of the polariton density matrix $\hat \rho(t)$, which describes the state of the cavity, to be given by the Lindblad quantum master equation \cite{Carmichael-book1,Breuer-book}
\begin{equation}\label{eq:Lindblad full}
\frac{d}{dt}\hat \rho(t) = -i[\hat{\mathcal{H}},\hat \rho(t)]+ \mathcal{L}_a\left[\hat\rho (t)\right]+ \mathcal{L}_\mu[\hat \rho(t)].
\end{equation}
The first term describes coherent polariton evolution as determined by the Holstein-Tavis-Cummings Hamiltonian  $\mathcal{H}$ [Eq. (\ref{eq:HTC})]. The term $\mathcal{L}_a\left[\hat\rho (t)\right]$ describes the dissipative evolution associated with the loss of cavity photons through the mirrors (cavity leakage), and the term $ \mathcal{L}_\mu[\hat \rho(t)]$ describes the dissipative dynamics associated with radiative dipole decay of the electronic degrees of freedom (fluorescence) into bound modes of the nanostructure. The Lindblad form of the dissipators read \cite{Carmichael-book1,Breuer-book}
\begin{eqnarray}
\mathcal{L}_a\left[\hat\rho (t)\right]&=&\frac{\kappa}{2}\left(2\hat a\hat \rho \hat a^\dagger -\{\hat a^\dagger\hat a,\hat \rho \}\right)\label{eq:Lindblad cavity}\\
 \mathcal{L}_\mu[\hat \rho(t)] &=& \sum_n \frac{\gamma_n}{2} \left(2\hat \sigma_n^{-}\hat \rho(t)\hat \sigma_n^{+}-\{\hat \sigma_n^{+}\hat \sigma_n^{-},\hat \rho(t)\}\right),\label{eq:Lindblad mu}
\end{eqnarray}
where $\kappa$ is the empty-cavity decay rate,  $\hat\sigma_n^-  \equiv\ket{g_n}\bra{e_n}$ is an electronic transition operator, with $\hat\sigma_n^+ =[\hat\sigma_n^-]^\dagger $, and $\{\hat A,\hat B\}$ denotes an anti-commutator. The single-emitter decay rate $\gamma_n$ can be estimated from the classical Green's function of the optical nanostructure \cite{Delga2014}, and is not necessarily equal to the fluorescence decay rate in free space \cite{Torma2015}. 

We implicitly assume in Eq. (\ref{eq:Lindblad mu}) that a typical distance between emitters in the ensemble $\langle (\x_n-\x_m)\rangle$ is much larger than the decay length of the equal-time two-point correlation function $\langle \hat{\mathcal{E}}_b^{(-)}(\x_n)\hat{\mathcal{E}}_b^{(+)}(\x_m) \rangle $. If this is not the case, then site non-diagonal terms in the dipole dissipator $ \mathcal{L}_\mu[\hat \rho(t)]$ should be taken into account, as well as coherent long-range interactions between emitters \cite{Zhu:2015}. We made a local-decay assumption here for simplicity.  

 We can assume that all molecules in the ensemble have  identical transition dipole moments, and thus fluorescence  rate $\gamma_e$, and also that the spectral density of the electromagnetic reservoir does not vary significantly in a frequency range on the order of $\sim \sqrt{N}\Omega$ centered at the cavity frequency $\omega_c$. The latter assumption is not accurate in the ultrastrong coupling regime of cavity-matter coupling \cite{Baudoin2011,Wang2016}. 

 Expanding the quantum master equation in Eq. (\ref{eq:Lindblad full}) in terms of an eigenbasis of the HTC Hamiltonian $\hat{\mathcal{H}}$, and only including terms that evolve at optical frequencies, we arrive at the polariton master equation
\begin{equation}\label{eq:LQME}
\frac{d}{dt}\hat \rho(t) = \sum_{ij} \frac{\gamma_{ij}}{2}\left(2 \ket{\epsilon_i}\bra{\epsilon_j} \hat \rho (t) \ket{\epsilon_j}\bra{\epsilon_i} - \left\{\ket{\epsilon_j}\bra{\epsilon_j},\hat \rho(t) \right\}  \right),
\end{equation}
where we use a notational convention in which eigenstates $\ket{\epsilon_i}$ belong to the ground electronic state manifold for the $i$-th vibrational configuration of the molecules, and states $\ket{\epsilon_j}$ represent polariton eigenstates (light-matter excitation). The transition frequencies $\omega_{ij}=\epsilon_j-\epsilon_i>0$ are in the optical regime of interest. The Lindblad (secular) form of the polariton master equation in Eq. (\ref{eq:LQME}) describes simple birth-death processes for the populations, as well as decay of polariton coherences $\rho_{ij} \equiv \bra{\epsilon_i}\hat \rho\ket{\epsilon_j}$ due to radiative relaxation.

From the Lindblad master equation above, we obtain a decay rate for the $j$-th polariton eigenstate $\Gamma_j = \sum_{i} \gamma_{ij}$ (see Appendix \ref{app:fluctuations})  given by
\begin{equation}\label{eq:Gammas}
\Gamma_{j} = \kappa\sum_i|\bra{\epsilon_i}\hat a\ket{\epsilon_j}|^2+N\gamma_e\sum_j|\bra{\epsilon_i}\hat J_-\ket{\epsilon_j}|^2, 
\end{equation}
where we introduced a size-normalized collective transition operator \mbox{$\hat J_- =\mu^{(+)}/\sqrt{N}$}. This is done for numerical convenience, given the impossibility of simulating microcavities with large $N\gg 10^4$. For a numerically accessible system with finite $N\sim 20$, a normalized dipole decay rate of order unity can be numerically computed and then scaled to realistic values of $N\gamma_e$ in Eq. (\ref{eq:Gammas}). This is done to  get an accurate estimate for the homogeneous spectral linewidth.

\section{Modelling Organic Cavity Spectroscopy}

\begin{figure}[t]
\includegraphics[width=0.47\textwidth]{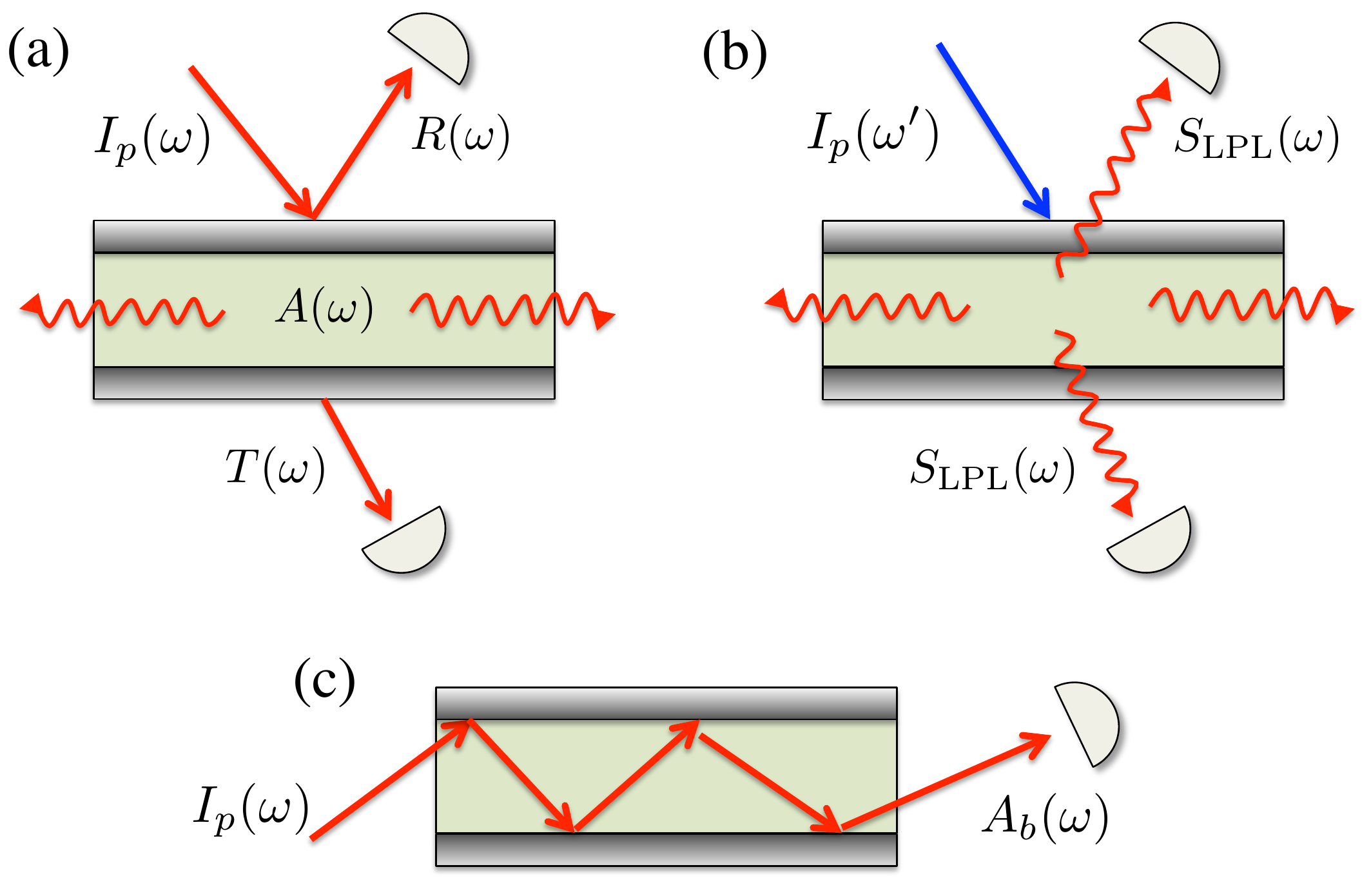}
\caption{Spectral signals. (a) Reflection $R$, transmission $T$, and absorption $A = 1-R-T$ of an external pump $I_p(\omega)$. Absorption is due to spontaneous emission into bound modes of the microcavity.  (b) Leakage photoluminescence $S_{\rm LPL}(\omega)$ following weak laser pumping at frequency $\omega'>\omega$. Bound mode photoluminescence \cite{Spano2015} is also shown. (c) Bound mode absorption spectra $A_b(\omega)$, for laser driving beyond cutoff for total internal reflection.}
\label{fig:definitions}
\end{figure}

Having described the structure and properties of dark vibronic polaritons of the $X$ and $Y$ types, our goal for the remainder of this article is to understand how dark vibronic polaritons manifest in the absorption and emission spectra of organic microcavities. We focus on the three types of measured signals: absorption through the mirrors $A(\omega)$, bound mode absorption $A_b(\omega)$, and photoluminescence through the mirrors or leakage photoluminescence $S_{\rm LPL}(\omega)$. In Fig. \ref{fig:definitions} we illustrate these types of spectral measurements. We do not consider  photoluminescence into bound modes of the nanostructure. This type of emission was treated in Ref. \cite{Spano2015} for J-aggregates in microcavities.

Conventional absorption through the mirrors $A(\omega)$ and bound mode absorption $A_b(\omega)$ differ in the way an external pump drives the cavity system. In the former case, a resonant pump laser transfers photons from a free-space electromagnetic mode (monochromatic laser) to the intracavity medium through the partially reflecting mirrors \cite{George2015-farad}. This photon transfer populates a polariton state $\ket{\epsilon_j}$, which can then decay either by converting an intracavity photon back to an external propagating photon, giving rise to reflection $R$ and transmission $T$, or by spontaneously emitting a photon into one of the bound modes of the nanostructure \cite{Matterson2001}. The latter decay channel leads to attenuation of the total reflected plus transmitted field, relative to the input photon flux. Since conservation of photon flux thus requires that $R+T+A = 1$ at all pump frequencies,  the conventional absorption spectra is defined as \cite{Schwartz2013}
\begin{equation}\label{eq:A}
A(\omega_p) = 1-R(\omega_p)-T(\omega_p),
\end{equation}
where $R(\omega_p)$ and $T(\omega_p)$ are the reflection and transmission spectra, normalized to the pump intensity at frequency $\omega_p$. This type of absorption spectra is thus an indirect way of measuring the material part of polaritons, i.e., the material polarization, because the magnitude of $A$ depends on the ability of a polariton to decay radiatively into bound modes, which is ultimately governed by the transition dipole operator $\hat \mu$. On the other hand, bound mode absorption $A_b(\omega)$, as well as fluorescence, can directly probe the material polarization $\hat \mu$, as the probe field is trapped inside the intracavity medium by total internal reflection \cite{Barnes1986,Matterson2001}. 

In general, non-radiative decay processes involving electronic degrees of freedom also contribute to $A(\omega)$, however the timescales for non-radiative molecular relaxation  $\tau_{\rm nr}\sim 1-10$ ps \cite{May-Kuhn} can be expected to be  slower than the radiative timescales $1/\kappa\sim 1/N\gamma_e\sim  10$ fs in low-$Q$ microcavities \cite{Hobson2002,Torma2015}. 

\subsection{Output Spectrum and Cavity Absorption}

The reflection spectra $R(\omega)$ and transmission spectra $T(\omega)$ are given by the spectrum of the electric field operator $\hat{\mathcal{E}}(\x,t)$ [$\sqrt{{\rm Hz}}$] on either external side of a microcavity driven by a pump photon flux,  denoted as $|\langle \hat F_1\rangle|^2$ [Hz], incoming on one side of the cavity. We show in Appendix \ref{app:input output}, using a Schr\"{o}dinger picture input-output formalism \cite{Carmichael-book1}, that if we denote by $\hat{\mathcal{E}}_1$ the field reflected on the same side as the incoming pump field, and by $\hat{\mathcal{E}}_2$  the field transmitted on the other side of the cavity, conservation of photon flux requires that 
\begin{equation}\label{eq:RT}
|\langle \hat{\mathcal{E}}_1\rangle |^2+|\langle \hat{\mathcal{E}}_2\rangle |^2 = |\langle \hat F_1\rangle|^2, 
\end{equation}
in steady state. This becomes the usual empty-cavity relation $R+T=1$ after normalization by the incoming flux. In general, Eq. (\ref{eq:RT}) needs to be integrated over all frequencies to obtain a relation in units of photon number.

For a cavity with an intracavity medium, as discussed above, polariton fluorescence into bound modes leads to attenuation of the reflected and transmitted fields. If we assume that a polariton can emit radiation at a single frequency, photon flux conservation reads
\begin{equation}\label{eq:BRT}
|\langle \hat{\mathcal{E}}_b\rangle |^2+|\langle \hat{\mathcal{E}}_1\rangle |^2+|\langle \hat{\mathcal{E}}_2\rangle |^2 = |\langle \hat F_1\rangle|^2,
\end{equation}
where $\hat{\mathcal{E}_b}$ is the bound electric field flux operator. Normalizing by the input flux gives Eq. (\ref{eq:A}), This relation shows that the through-mirror absorption spectra $A(\omega) = 1-R(\omega)-T(\omega)$ is determined by the steady-state spectrum of the the bound electric field $S_b(\omega) \propto \int d\tau \langle \hat{\mathcal{E}_b}^\dagger(\tau)\hat{\mathcal{E}}_b(0)\rangle {\rm e}^{i\omega\tau} $, which as shown in Appendix \ref{app:absorption}, is proportional to the bound fluorescence spectra
\begin{equation}
S_b(\omega) \propto \int_0^\infty\, d\tau\, \langle \hat \mu^{(-)}(\tau)\hat \mu^{(+)}(0)\rangle\, {\rm e}^{i\omega\tau},
\end{equation}
where $\hat \mu^{(+)}$ is defined in Eq. (\ref{eq:mu}). Integrating the fluorescence spectra over all emission frequencies gives the total number of photons emitted into the bound modes, after pumping the cavity with a monochromatic field at frequency $\omega_p$. Therefore, the value of the absorption coefficient $A$ at the pump frequency $\omega_p$ is given by
\begin{equation}\label{eq:absorption def}
A(\omega_p) = \int_0^\infty d\omega\, \int_0^\infty d\tau\, \langle \hat \mu^{(-)}(\tau)\hat \mu^{(+)}(0)\rangle\, {\rm e}^{i\omega\tau},
\end{equation}
up to a proportionality factor that depends on the geometry of the cavity. 

In Appendix \ref{app:weak absorption}, we develop a model to evaluate the spectrum of the dipole fluctuations in Eq. (\ref{eq:absorption def}), which determine the absorption of a weak pump field at frequency $\omega_p$. The external periodic driving is considered as a perturbation to the Holstein-Tavis-Cummings Hamiltonian $\hat{\mathcal{H}}$, which defines a time-dependent driven HTC model of the form $\hat H(t) = \hat{\mathcal{H}}+\hat V_p(t)$, where 
\begin{equation}\label{eq:Vp}
\hat V_{\rm p}(t) = \Omega_p\left(\hat a^\dagger\,{\rm e}^{-i\omega_p t}+\hat a\,{\rm e}^{i\omega_p t}\right),
\end{equation}
is the periodic driving term with frequency $\omega_p$ and amplitude $\Omega_p\ll \sqrt{N}\Omega\ll \omega_c$. We obtain the steady-state polariton population $\rho_j$ of the system in the presence the external driving, to second order in $\Omega_p$. The dipole fluctuations are evaluated within the no-quantum-jump (NQJ) approximation, which we introduce to simplify the quantum regression formula. After performing the time and frequency integrations in Eq. (\ref{eq:absorption def}), we obtain the expression
 \begin{equation} \label{eq:absorption}
 A(\omega_p) =\pi |\Omega_p|^2\sum_j\frac{|\langle G|\hat a|\epsilon_j\rangle|^2 (\kappa_{Gj}/\Gamma_j)}{(\omega_p-\omega_{Gj})^2+\kappa_{Gj}^2}F_j,
 \end{equation}
where $\kappa_{Gj}$ is the decay rate for the optical coherence $\rho_{Gj} \equiv \bra{G}\hat \rho\ket{\epsilon_j}$, which we can allow to account for non-radiative relaxation processes as well (details in Appendix \ref{app:weak absorption}). We have also defined the total dipole emission strength of the $j$-th polariton eigenstate as
\begin{equation}\label{eq:Fj}
F_j = \sum_i|\langle\epsilon_i|\hat \mu^{(+)}| \epsilon_j\rangle|^2.
\end{equation}
 Equation (\ref{eq:absorption}) shows that if $F_j = 0$ for a given polariton eigenstate $\ket{\epsilon_j}$, there is no resonant absorption at that polariton frequency either.  In other words, polaritons that fluorescence poorly into bound modes of the nanostructure, cannot attenuate the reflected and transmitted fields efficiently. On the other hand, even when a polariton state $\ket{\epsilon_j}$ can have a strong bound fluorescence, it cannot attenuate the driving field efficiently when its vibrationless photonic component $\langle G|\hat a|\epsilon_j\rangle $ is negligible or the pump field is far detuned from $\omega_{Gj}$, which suppresses its  stationary polariton population $\rho_j$.

\subsection{Cavity Photoluminescence}
\label{sec:LPL}

Leakage photoluminescence (LPL) is detected on either side of an organic cavity, as illustrated in Fig. \ref{fig:definitions}. The LPL spectra $S_{\rm LPL}(\omega)$ is thus directly proportional to the spectrum of the external field operators $\hat{\mathcal{E}}_1(t)$ and $\hat{\mathcal{E}}_2(t)$, at the reflection or the transmission side of the cavity, respectively. As discussed in Appendix \ref{app:input output}, for a classical (coherent state) input field that is uncorrelated with the intracavity field operator $\hat a(t)$, i.e., $\langle \hat F_1(t)\hat a(t')\rangle \rightarrow 0$, the external field spectrum is directly proportional to the spectrum of the intracavity field fluctuations. Therefore, the stationary leakage PL spectra can be computed as
\begin{equation}\label{eq:LPL def}
S_{\rm LPL}(\omega) = \int_0^\infty d\tau \, \langle \hat a^\dagger(\tau) \hat a(0)\rangle \, {\rm e}^{i\omega\tau},
\end{equation}
up to a geometry-dependent proportionality factor. The dependence on the pump frequency enters in the intracavity field autocorrelation dynamics, which we compute from the dissipative dynamics of organic polaritons, together with the quantum regression formula.  

We show in Appendix \ref{app:fluctuations} that the spectrum of the stationary cavity fluctuations in Eq. (\ref{eq:LPL def}) can be easily evaluated within the NQJ approximation. The resulting expression for the LPL spectra reads
\begin{equation}\label{eq:S LPL}
S_{\rm LPL}(\omega) = \sum_{ij}\rho_j\,|\langle\epsilon_i| \hat a| \epsilon_j\rangle |^2 \frac{(\Gamma_{j}/2)}{(\omega-\omega_{ij})^2+(\Gamma_{j}/2)^2},
\end{equation}
where $\rho_j$ is the steady-state population of the $j$-th HTC polariton eigenstate $\ket{\epsilon_j}$,  $\Gamma_j$ is the polariton decay rate from Eq. (\ref{eq:Gammas}), and $\omega_{ij} = \epsilon_j-\epsilon_i>0$ is the transition frequency for the optical transition $\ket{\epsilon_j}\rightarrow \ket{\epsilon_i}$. 

We compute the  LPL emission spectra below by assuming a  stationary polariton population of the form $\rho_j=1/Md_j$, where $d_j$ is the degeneracy of the $j$-th HTC eigenvalue and $M=\sum_j d_j$ is the number of polariton states considered. For a single molecule $d_j=1$ for all $j$, but for $N\geq 2$ this is not the case.

\section{Absorption and Emission of Organic Cavities}

We now apply the quantum formalism developed in the previous sections to discuss the absorption and photoluminescence spectroscopy of organic microcavities. We discuss first the case of a single molecule in a cavity ($N=1$), then a molecular dimer ($N=2$), and finally a molecular ensemble with large $N$. For the ensemble case, we compare the simulated  emission spectra with experimental results from the literature.

\begin{figure}[h]
\includegraphics[width=0.4\textwidth]{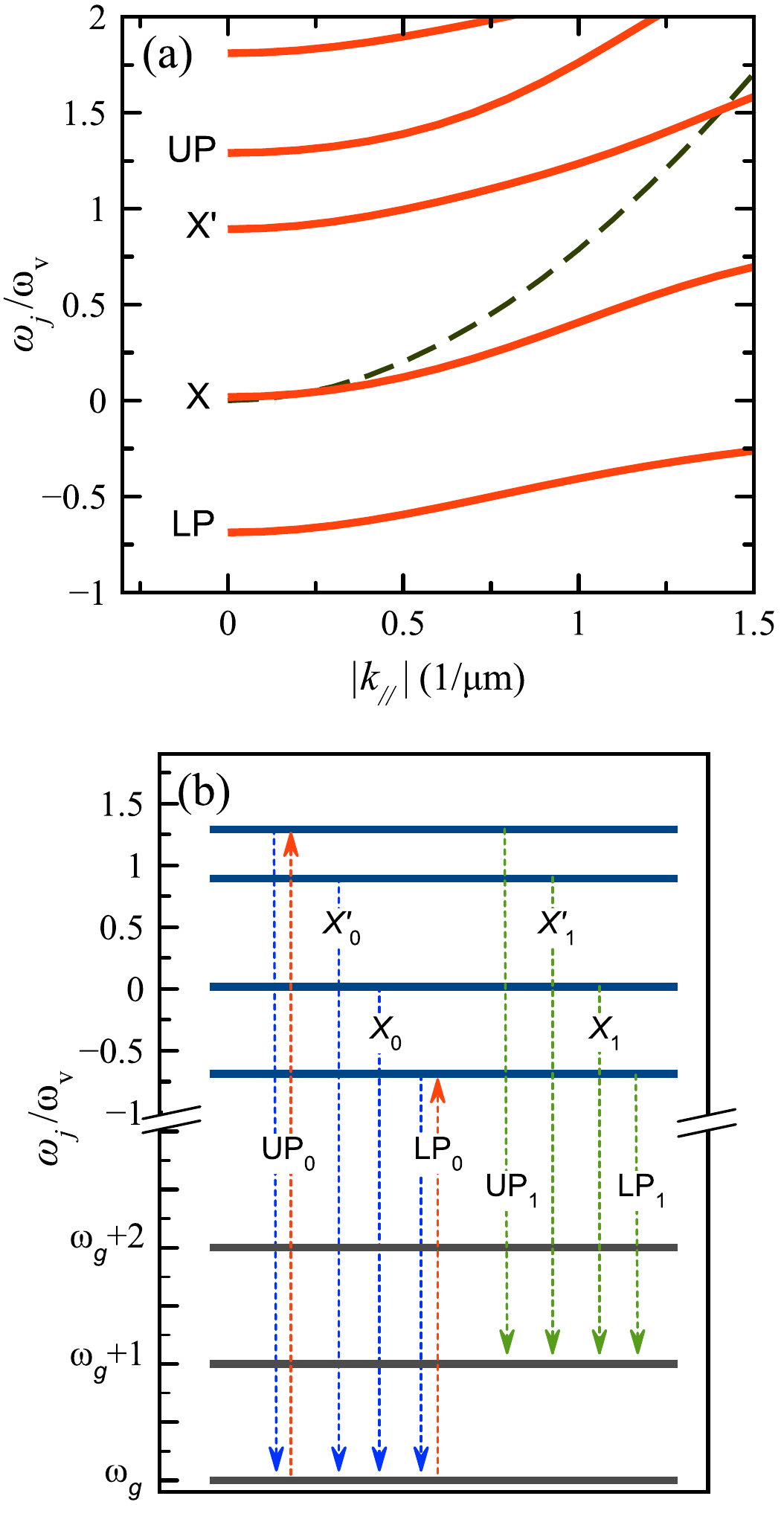}
\caption{ Single molecule dispersion and allowed transitions. (a) Polariton dispersion for Rabi coupling $\Omega = 1.68 \,\omega_{\rm v}$. For all considered in-plane wave vectors $k_\parallel$, the polariton branches have at least 10 \% photonic character.  (b) Energy level diagram with the arrows indicating allowed transitions from the bottom of the polariton branches in panel (a). Transitions are denoted as $P_
\nu$, from a polariton state $P=\ket{\epsilon_j}$ at frequency $\omega_j$, into a state of the ground manifold with $\nu$ vibrational quanta  $\ket{g,\nu,0_c}$. Upward arrows are dipole allowed transitions, relevant in bound absorption experiments, for  $\omega_g$ being the absolute ground state energy (vertical axes not on scale). In both panels we set $\lambda^2 = 1$. $\omega_{\rm v}$ is the intramolecular vibration frequency. }
\label{fig:single dispersion}
\end{figure}

\subsection{Single Molecule}
 
For a single emitter, the critical Rabi coupling for which the zero-eigenvalue equation (\ref{eq:HX0}) can be solved is $\Omega=1.68\,\omega_{\rm v}$ for $\lambda^2=1$, at normal incidence ($k_\parallel = 0$). For this choice of Rabi frequency, we show the polariton dispersion in Fig. \ref{fig:single dispersion}(a). The dispersion of the empty cavity mode $\omega_c(k_\parallel)$ is shown for comparison (dashed line), where $k_\parallel$ is the in-plane wave vector of the cavity mode. For all values of $k_\parallel$ shown, the HTC polariton eigenstates have at least 10$\%$ photonic character. The zero-energy dark vibronic polariton state $\ket{X}$, which is dispersive in our model, appears in the middle region between the lower (LP) and upper (UP) polariton branches. A second $X$-type polariton branch, which we denote as $\ket{X'}$, occurs near the upper polariton branch. 

Figure \ref{fig:single dispersion}(b) is a diagram of the allowed transitions from the bottom of each of the polariton branches shown in Fig. \ref{fig:single dispersion}(a), to the lowest three vibrational states of the ground state manifold.  Arrows indicate allowed transitions between polariton states $\ket{\epsilon_j}$, with frequencies $\omega_j$, and states in the ground manifold $\ket{g,\nu,0_c}$ having $\nu\leq 1$ vibrational quanta.  Upward arrows are dipole allowed transitions, relevant in bound absorption experiments \cite{Barnes1998,Matterson2001}. The energy of the absolute ground state $\ket{g,0,0_c}$ is denoted as $\omega_g$. Emission events are dissipative in the sense that the material is projected into the state $\ket{g,\nu,0_c}$ after photon loss through the cavity mirrors (leakage), which can be represented by the mapping
\begin{equation}\label{eq:PL map}
\hat a\ket{\epsilon_j}\rightarrow \ket{g,\nu,0_c}+\hbar\omega.
\end{equation}
The radiated photon is detected at the frequency 
\begin{equation}
\omega = \omega_j-\nu\,\omega_{\rm v},
\end{equation}
so that leakage photoluminescence maps the polariton dispersion $\omega_j(k_\parallel)$ only for vibrationless transitions ($\nu=0$). For convenience, we denote downward transitions by  $P_\nu$, where $P$ labels the emitting polariton state $\ket{\epsilon_j}$ and $\nu$ the number of vibrational quanta in the ground manifold after photon leakage. Dissipative transition of the form $P_{\nu>0}$  are ignored in existing quasi-particle theories of organic microcavities \cite{Agranovich2003,Litinskaya2004,Litinskaya2006,Michetti2008,Mazza2009,Cwik2016}.

\begin{figure}[h]
\includegraphics[width=0.4\textwidth]{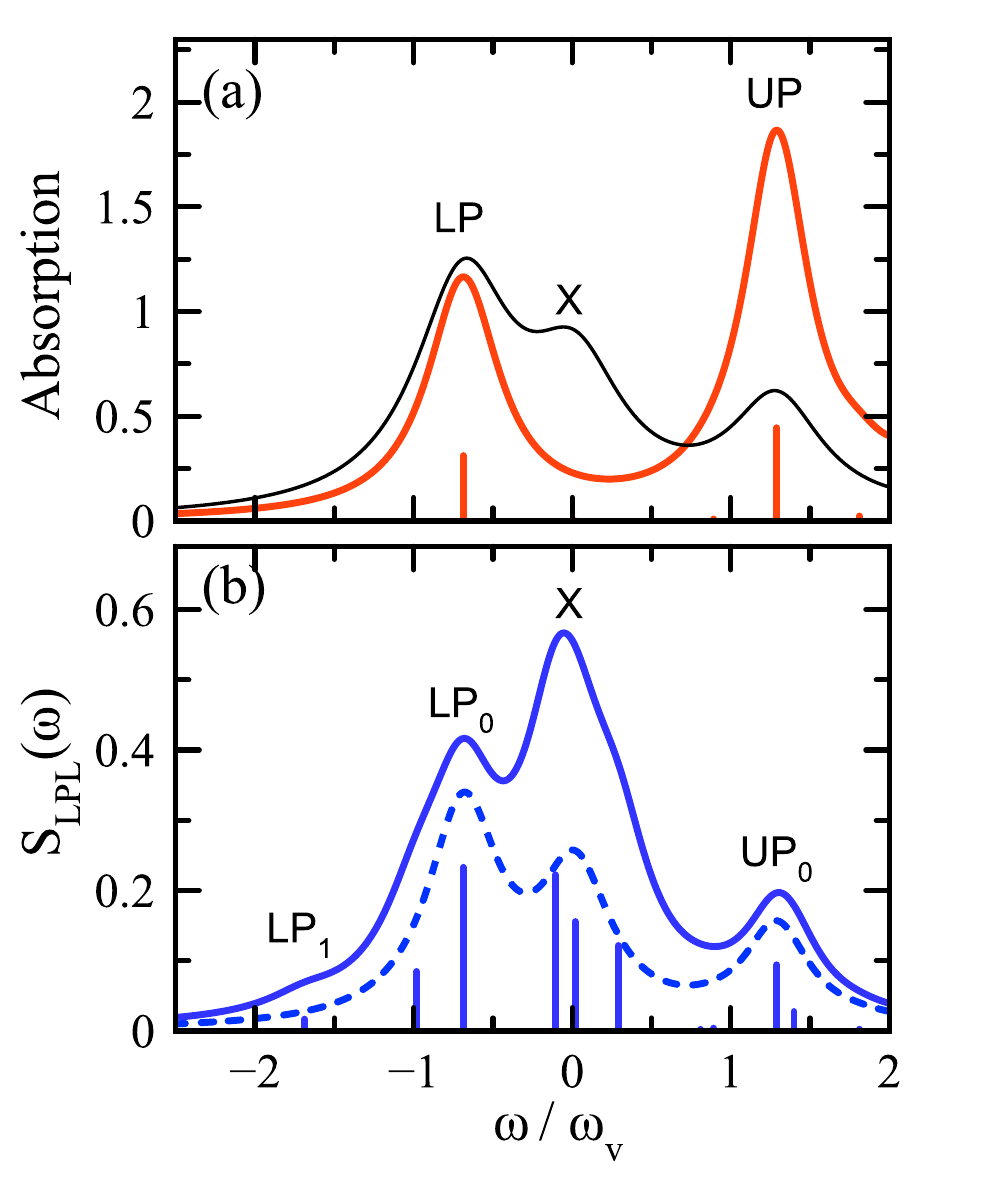}
\caption{{Cavity absorption and emission for a single molecule ($N=1$)}. (a) Conventional absorption $A = 1-R-T$ (black line) and bound absorption spectra (red line). The lower polariton (LP) and upper polariton (UP) peaks are present in both types of absorption signals. The zero-energy dark vibronic polariton $\ket{X}$ is not present in bound absorption. (b) Leakage photoluminescence spectra (LPL) for the same parameters as in panel (a). Peak labels have the form $P_\nu$, where $P$ refers to the initial polariton state and $\nu$ to the vibrational quantum number in the final state of the transition. Dashed and solid lines include transitions with up to $\nu=0$ and $\nu=1$ vibrational quanta, respectively. Vertical bars indicate the position and relative strength of the peak maxima. In both panels we set $\Omega=1.68\, \omega_{\rm v}$, $\kappa/\omega_{\rm v} = 0.9$, $\gamma_e/\omega_{\rm v}=0.2$, and a non-radiative decay rate $\gamma_{\rm nr}/\omega_{\rm v}= 0.1$, for $\lambda^2 = 1$ and $\omega_c=\omega_{00}$ at normal incidence. $\omega_{\rm v}$ is the intramolecular vibration frequency.}
\label{fig:single spectra}
\end{figure}

We show in Fig. \ref{fig:single spectra} the simulated absorption and leakage PL emission spectra for a single molecule in a microcavity. Such systems can now be realized experimentally using localized plasmonic fields \cite{Chikkaraddy:2016aa}. The bound absorption spectra is computed from the dipole autocorrelation function as discussed in Appendix \ref{app:fluctuations}, assuming that only the absolute ground state $\ket{G}$ is populated. The lower polariton and upper polariton form a well-defined bound absorption doublet, for the transitions indicated with upper arrows in Fig. \ref{fig:single dispersion}(b). While the dipole oscillator strength is exactly zero for the $\ket{X}$ state, making it invisible in bound absorption, its non-vanishing photonic component $\langle G|\hat a|X\rangle $ gives a peak in conventional absorption.

The leakage photoluminescence (LPL) spectra, shown in Fig. \ref{fig:single spectra}, illustrates the importance of dissipative transitions of the form in Eq. (\ref{eq:PL map}) with $\nu\geq 1$. The spectra is calculated as described in Sec. \ref{sec:LPL}. If we only take into account emission events into the vibrationless ground state of the cavity (dashed line) we obtain an photoluminescence spectra that roughly maps the conventional absorption spectra $A(\omega)$ in terms of its relative peak intensities. For example, the ratio between the LP and X peaks in conventional absorption is similar to their ratio in PL emission. However, if we include emission processes that project the system into a state with up to $\nu=1$ quantum of vibration, the emission peak near the bare molecular frequency is enhanced relative the lower polariton peak. Allowed transitions of this kind are shown in Fig. \ref{fig:single dispersion}(b).

Leakage transitions from polariton eigenstates $\ket{\epsilon_j}$ that leave the material with $\nu\geq 2$ vibrational quanta are also possible, and in general contribute to the strength of the emission bands. In this work, we assume that only polaritons that are slightly higher in energy than the conventional upper polariton state are populated and can contribute to the LPL spectra. Under the assumption of uniform polariton population adopted here,  photon leakage into a ground state with $\nu\geq 2$ quanta do not qualitatively change the emission spectra within the conventional lower and upper polariton doublet. We therefore omit those transitions from the discussion for clarity. The situation is different when a more localized polariton population is assumed \cite{Herrera2016-PRL}. In this case, emission at the lower polariton frequency can be dominated by leakage transitions with $\nu=2$ from two-particle polariton eigenstates in the vicinity the upper polariton frequency, which is approximately two vibrational quanta above the lower polariton state for $\sqrt{N}\Omega/\omega_{\rm v}\approx 2$. 

\subsection{Molecular Dimer}

\begin{figure}[t]
\includegraphics[width=0.4\textwidth]{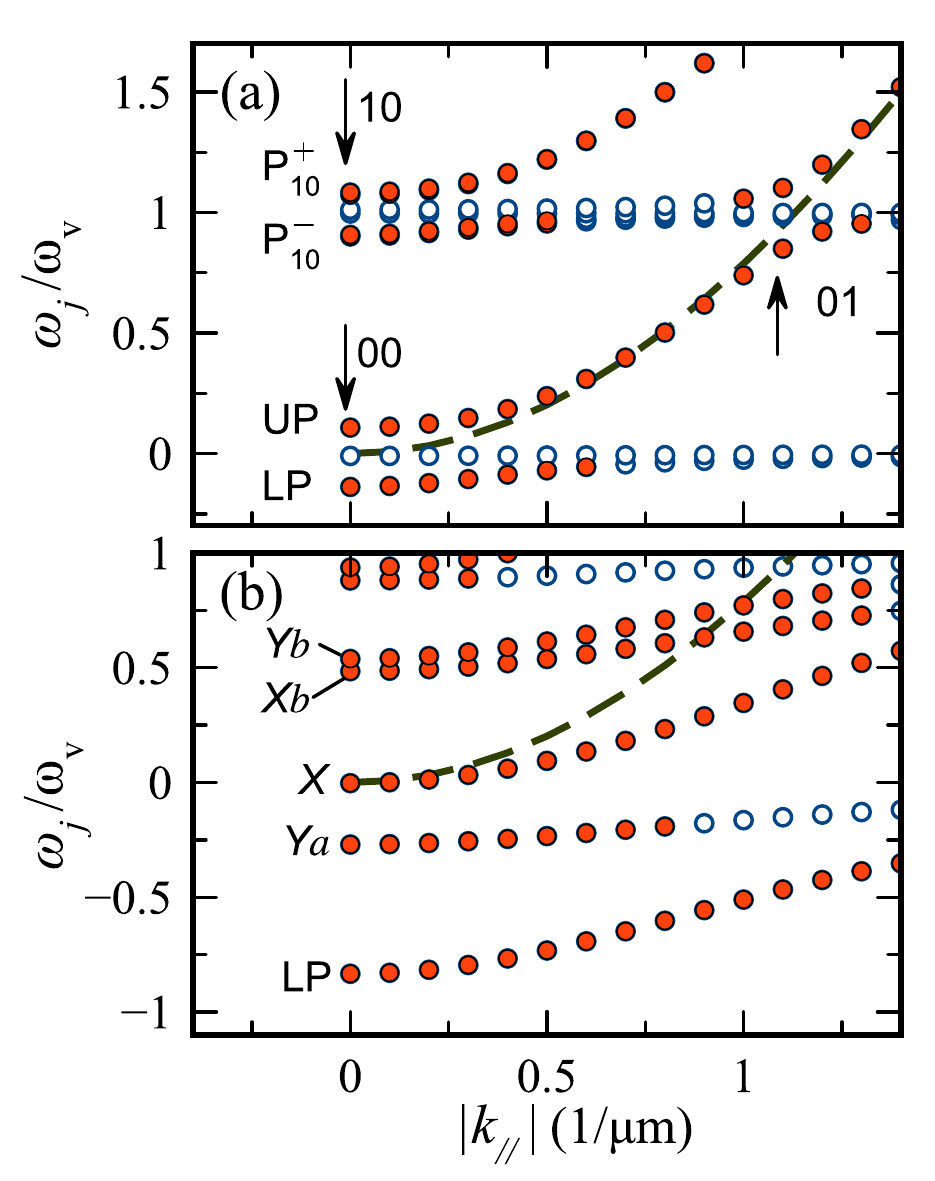}
\caption{{Polariton dispersion for a dimer in a cavity ($N=2$)}. (a) Dispersion for Rabi coupling $\sqrt{N} \Omega=0.4\,\omega_{\rm v}$. The states P$_{10}^\pm$ are two-particle polaritons eigenstates in this regime. Arrows indicate the location of an avoided crossing labelled by the pair of quantum numbers $(\nu\,\tilde \nu')$, where $\nu$ refers to the vibrational quantum number of the photon state and $\tilde \nu$ to the material vibronic excitation involved in polariton formation. The bare cavity dispersion is also shown (dashed line). Polariton states with at least 10\% photonic character are shown with filled circles. Open circles indicate polariton states that have predominantly material character. (b) Dispersion of the lowest polariton states for Rabi coupling $\sqrt{N} \Omega=2.0\,\omega_{\rm v}$. We set $\lambda^2 = 1$ in both panels. $\omega_{\rm v}$ is the intramolecular vibration frequency.}
\label{fig:dimer dispersion}
\end{figure}

The dimer $(N=2)$ is the simplest scenario where two-particle states need to be taken into account in order to describe the spectra. Single-particle and two-particle material and polariton states are described in Section \ref{sec:DVP}. We first consider the polariton dispersion for a dimer in two different Rabi coupling regimes. For relatively small Rabi frequencies $\sqrt{N} \Omega/\omega_{\rm v}\ll 1$, polaritons that arise from the lowest vibronic transition of the molecule ($\tilde \nu = 0$) have mostly single-particle character, forming the usual lower polariton and upper polariton branches. In this Rabi coupling regime, two-particle states that originate from higher vibronic bands also lead to polariton splitting at normal incidence ($k_\parallel =0$) \cite{Spano2015}.  We show the corresponding polariton dispersion in Fig. \ref{fig:dimer dispersion}(a), which displays the usual anti-crossings between vibronic transitions with $\tilde \nu =0$ and $\tilde \nu=1$ vibrational quanta, with the vibrationless cavity photon dispersion (dashed line). As we discussed in Section \ref{sec:DVP}, the Holstein-Tavis-Cummings (HTC) model also allows the formation of the two-particle {\it polariton} states $\ket{P_{\nu\tilde \nu'}^\pm,\beta}$  [Eq. (\ref{eq:TP polariton})] at normal incidence, a fact largely ignored in previous works on vibronic polaritons \cite{Michetti2008,Mazza2009}. The  polariton splittings associated with two-particle polaritons are smaller than for  conventional (single-particle) lower and upper polariton splitting by a factor $\sqrt{1-(1/N)}$, due to the saturation nonlinearity involved in  vibronic-vibrational configurations (see Fig. \ref{fig:SP-TP}). While for large $N\gg 1$, the difference between single particle and two-particle polariton splittings can be negligible,  for small size systems with $N\sim 1$ \cite{Chikkaraddy:2016aa}, it may be possible to resolve the difference in polariton splittings between single-particle and multi-particle vibronic polaritons.

\begin{figure}[t]
\includegraphics[width=0.4\textwidth]{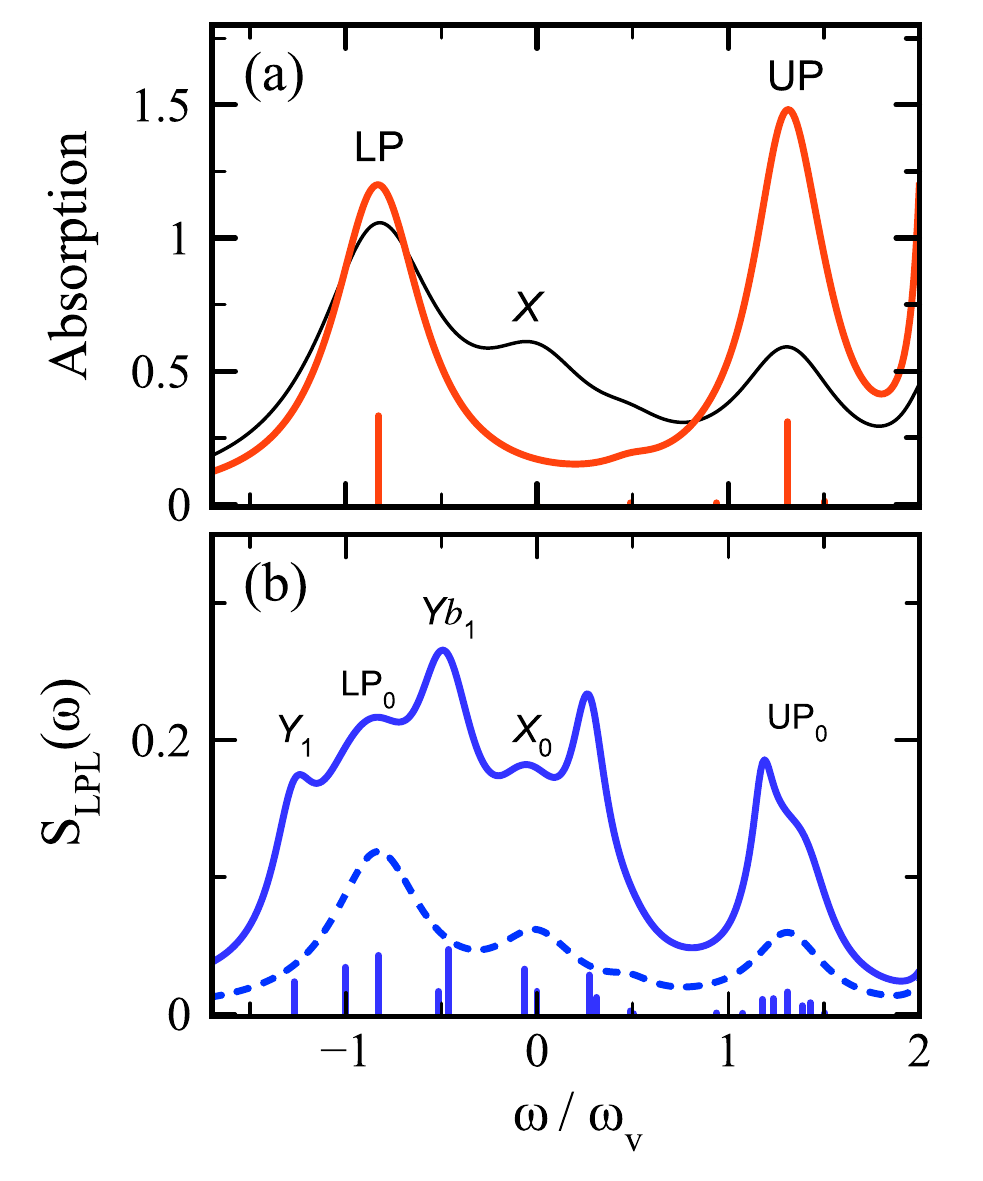}
\caption{{Cavity absorption and emission for a dimer in a cavity ($N=2$)}. (a) Conventional absorption $A = 1-R-T$ (black line) and bound absorption spectra (red line). The lower polariton (LP) and upper polariton (UP) peaks are present in both types of absorption signals. The zero-energy dark vibronic polariton $\ket{X}$ is not present in bound absorption. (b) Leakage photoluminescence spectra (LPL) for the same parameters as in panel (a). Peak notation follows Fig. \ref{fig:single spectra}. Dashed and solid lines include transitions with up to $\nu=0$ and $\nu=1$ vibrational quanta, respectively, in the ground manifold. Vertical bars indicate the position and relative strength of the peak maxima. In both panels we set $\Omega=2.0\, \omega_{\rm v}$, $\kappa/\omega_{\rm v} = 0.9$, $N\gamma_e/\omega_{\rm v}=0.2$, and a non-radiative decay rate $\gamma_{\rm nr}/\omega_{\rm v}= 0.1$, for $\lambda^2 = 1$ and $\omega_c=\omega_{00}$ at normal incidence. $\omega_{\rm v}$ is the intramolecular vibration frequency.}
\label{fig:dimer spectra}
\end{figure}
	
	For Rabi frequencies closer to the critical value for the formation of the zero-energy $\ket{X}$ state,  i.e., $\sqrt{N} \Omega/\omega_{\rm v}\approx 2$ (see Fig. \ref{fig:HTC params}), all eigenstates of the HTC model acquire a more significant contribution from two-particle material states. We show in Fig. \ref{fig:dimer dispersion}(b) the polariton dispersion for this coupling regime.  At normal incidence, dark vibronic polaritons of the $X$ type (see Section \ref{sec:DVP}) are present in the middle region between the conventional lower and upper polariton branches. We label the dark vibronic polariton at zero energy by $X$, and the vibronic polariton that is blue shifted from the bare electronic frequency by  $Xb$ ($\omega_{Xb}\approx 0.5\, \omega_{\rm v}$).  The spectrum also features additional branches in the middle region between the conventional lower and upper polariton states. These correspond to dark vibronic polaritons of the $Y$ type (Sec. \ref{sec:DVP}), and are labelled as ${ Ya}$ and  ${ Yb}$. As discussed in Section \ref{sec:Y type}, these states are dominated by two-particle material states, and can be weakly dispersive while still having a significant photon character.
	
In Fig. \ref{fig:dimer spectra}, we show the computed absorption and leakage photoluminescence (LPL) spectra for the bottom of the polariton branches in Fig. \ref{fig:dimer dispersion}. The conventional and bound absorption spectra are qualitatively similar to the single molecule case (Fig. \ref{fig:single spectra}), with both $X$-type dark vibronic polaritons being visible in conventional through-mirror absorption, but only the state $Xb$ being weakly visible in the bound absorption spectra, due to an incomplete destructive interference effect (Sec. \ref{sec:X type}). Unlike the single molecule case, the LPL spectra for a dimer has a much richer structure when considering photon leakage transitions from dark vibronic polaritons of the $Y$ type that leave the material with up to $\nu=1$ vibrational excitation. Such $Y$ states are invisible in both types of absorption signals. Most notable is the appearance of multiple emission peaks in the vicinity of the conventional lower polariton peak (${\rm LP}_0$), which results in a broad unresolved structure for realistic linewidths. However, when  considering only vibrationless photon leakage transitions ($\nu=0$), the LPL peak structure then simply resembles the conventional absorption spectra.

\subsection{Molecular Ensembles}

We finally consider the spectra of large molecular ensembles in a cavity, in an effort to provide insight on the several reported features of the absorption and emission spectra \cite{Hobson2002,Coles2011,Virgili2011,Schwartz2013,George2015-farad} for which quasi-particle theories \cite{Agranovich2003,Litinskaya2004,Litinskaya2006,Michetti2008,Mazza2009,Cwik2016} do not offer a consistent interpretation. We compute the dispersion for a system with $N=10$  molecules and the spectra for  $N=20$, which are representative values of the large $N$ limit \cite{Zeb2016}.

\begin{figure}[t]
\includegraphics[width=0.40\textwidth]{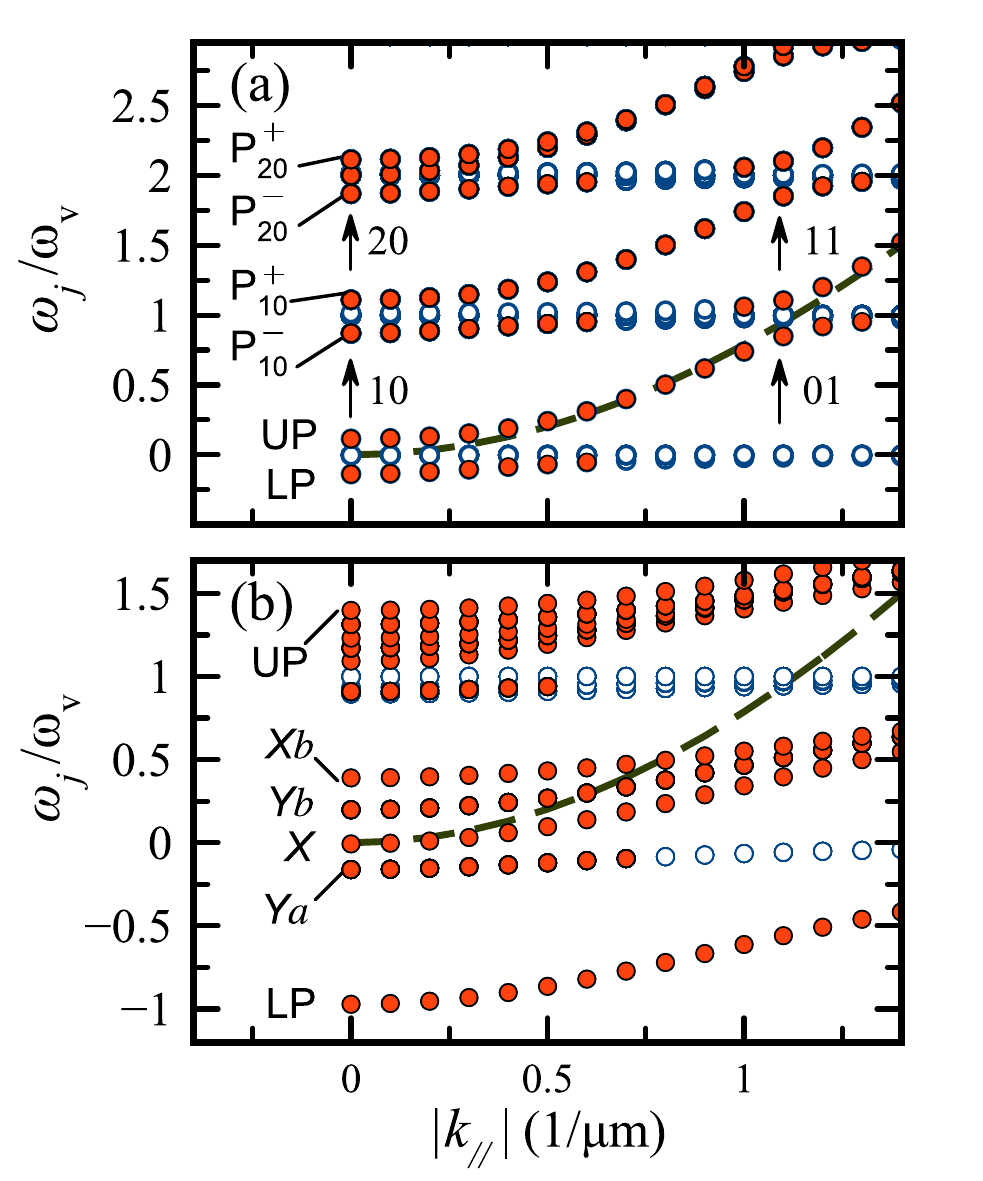}
\caption{Polariton dispersion for $N=10$ molecules in a cavity with Rabi coupling $\sqrt{N} \Omega=0.4\,\omega_{\rm v}$ at normal incidence. The states P$_{ \nu\tilde\nu}^\pm$ are two-particle polaritons eigenstates in this regime. Avoided crossing are labelled by the pair of quantum numbers $(\nu\,\tilde \nu')$, where $\nu$ refers to the vibrational quantum number of the photon state and $\tilde \nu$ to the material vibronic excitation involved in polariton formation. The bare cavity dispersion is also shown (dashed line). Polariton states with at least 10\% photonic character are shown with filled circles. Open circles indicate polariton states that have predominantly material character. (b) Dispersion of the lowest polariton states for Rabi coupling $\sqrt{N} \Omega=2.4\,\omega_{\rm v}$We set $\lambda^2 = 1$. $\omega_{\rm v}$ is the intramolecular vibration frequency.}
\label{fig:ten dispersion}
\end{figure}

As we discussed for the dimer case, two-particle vibronic polaritons $\ket{P_{\nu\tilde \nu}^\pm,\beta}$ are associated with Rabi splittings at frequencies greater or equal than one vibrational quanta above the conventional lower polariton splitting, as Fig. \ref{fig:ten dispersion}(a) shows. The ratio  between the two-particle Rabi splittings  and the conventional lower and upper polariton splitting differs from unity by a factor of order $1/N$, which is negligible for large $N$. The two-particle polariton branch labelled $P_{10}^-$ in Fig. \ref{fig:ten dispersion}(a) becomes particularly important for the interpretation of the emission spectra at larger Rabi frequencies. The diabatic energy of the associated $N$-fold degenerate two-particle polariton states $\ket{\alpha_0\beta,\tilde 0\,1,-}$ [see Eq. (\ref{eq:TP polariton})] is in the vicinity of the bare electronic frequency ($\omega_j=0$) for Rabi couplings $\sqrt{N}\Omega\approx 2\,\omega_{\rm v}$, and can therefore admix with single-particle diabatic material states $\ket{\beta, \tilde 0,0_c}$, which is also in that frequency region, forming a set of dark vibronic polaritons of the $Y$ type for $\beta\neq \alpha_0$. These dark states are $(N-1)$-fold degenerate, and have a significant $\nu=1$ photonic component. Photon leakage through the mirrors would thus produce light  near the conventional lower polariton frequency $\omega_{\rm LP}$, which is roughly one quantum of vibration below the bare molecular frequency for the values of $\sqrt{N}\Omega$ and $\lambda^2$. 
Moreover, contributions from the two-particle diabatic polariton state labelled  $P_{2\tilde 0}^{-}$ in Fig. \ref{fig:ten dispersion}(a), would admix with the conventional (single-particle) upper polariton state for large Rabi couplings to produce a set of $Y$-type dark vibronic polaritons at a frequency $\omega_j\approx \omega_{\rm v}$,  near the conventional upper polariton frequency. The structure of dark vibronic polaritons of the $Y$ type is discussed in Sec. \ref{sec:DVP}.

\begin{figure}[t]
\includegraphics[width=0.40\textwidth]{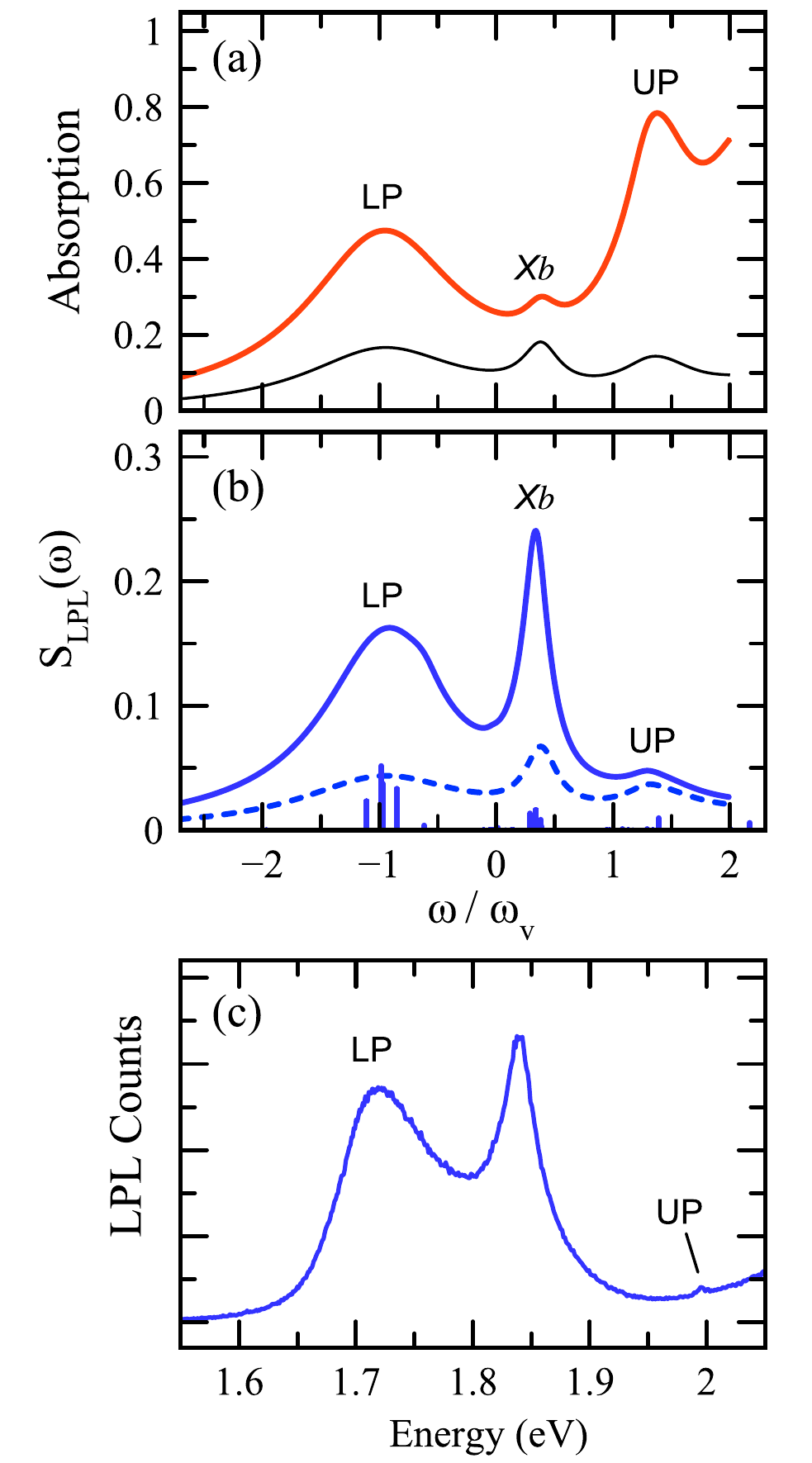}
\caption{{Cavity absorption and emission for a microcavity with $N=20$ emitters}. (a) Conventional absorption $A=1-R-T$ (black line) and bound absorption spectra (red line). (b) Normal incidence LPL spectra including transitions with up to $\nu=0$ and $\nu=1$ vibrational quanta in the ground manifold, shown in solid and dashed lines, respectively. Vertical bars indicate the position and relative strength of the peak maxima. We set $\sqrt{N}\Omega=2.4 \,\omega_{\rm v}$, $\lambda^2 = 1$, $\kappa/\omega_{\rm v} = 0.5$, and $N\gamma_e/\omega_{\rm v}=4$. $\omega_{\rm v}$ is the intramolecular vibration frequency. (c) Experimental LPL spectra obtained by Hobson {\it et al.} \cite{Hobson2002}, for cyanine dye J-aggregates in a low-Q microcavity (with permission). The cavity was pumped with a narrow continuous-wave laser at 2.08 eV.}
\label{fig:twenty spectra}
\end{figure}

As the polariton dispersion in Fig. \ref{fig:ten dispersion}(b) shows, for Rabi couplings $\sqrt{N}\Omega\approx 2\,\omega_{\rm v}$ there are several $Y$-type dark vibronic polaritons deriving from $\ket{P_{2\tilde 0}^-,\beta}$ that have energies near the upper polariton ($\omega_j\approx \omega_{\rm v}$). Figure \ref{fig:twenty spectra}(a) shows that dark vibronic polariton states of the $Y$ type are invisible in both types of absorption spectra, as for the dimer case. Such $Y$ states can nevertheless contribute to photoluminescence peak near the conventional lower polariton frequency $(\omega_j\approx -\omega_{\rm v})$, when the system is projected into a state with two vibrational quanta upon photon leakage, and also to the peak near the bare electronic frequency $(\omega_j\approx 0)$ when emitting light that leaves the system with one quantum of vibration. Similarly, dark vibronic polaritons of the $X$ and $Y$ types near the bare electronic frequency, i.e, branches labelled {\it Ya}, {\it X}, {\it Yb}, and {\it Xb} in Fig. \ref{fig:ten dispersion}(b), also contribute to PL emission near the lower polariton frequency, when projecting the system into a state with one vibrational excitation.


We show in Fig. \ref{fig:twenty spectra}(b) the calculated photoluminescence spectra (LPL) representative of molecular ensembles in organic cavities. This should be compared with the experimental PL spectra in Fig. \ref{fig:twenty spectra}(c), obtained by Hobson  {\it et al.}\cite{Hobson2002} for an ensemble of cyanine dye J-aggregates in a metallic microcavity at room temperature. The model spectra qualitatively reproduces the observed emission features, and is also consistent with other recent measurements on a different type of J-aggregate (TDBC)  \cite{Schwartz2013}. Similar to the single-molecule and dimer cases, if we ignore emission from $Y$-type dark vibronic polaritons that leave the material with $\nu\geq 1$ vibrational quanta upon photon leakage, the relative peak strengths differ significantly from the experimentally observed ratios. 

We also note the relative simplicity of the spectra for large $N$ relative to the dimer case. The multiple emission lines near the lower polariton frequency $\omega_{\rm LP}$ in Fig. \ref{fig:dimer spectra}, for the dimer case, merge into a single broad emission band whose peak maximum is blue-shifted relative to $\omega_{\rm LP}$ for an  ensemble. For the chosen model parameters, the shift is due to emission from the ($N-1$)-fold degenerate dark vibronic polariton labelled {\it Yb} in Fig. \ref{fig:ten dispersion}(b). For $N=20$  this state has energy  $\omega_Y = 0.148 \,\omega_{\rm v}$, and can emit light with frequency $\omega_Y-\omega_{\rm v} \approx -0.85\,\omega_{\rm v}$ upon mirror leakage, which is blue shifted from the lower polariton frequency $\omega_{\rm LP}$ by $\delta_{\rm LP}= 0.12\,\omega_{\rm v}$. This value of $\delta_{\rm LP}$ corresponds to $20$ meV for a vinyl stretch ($\omega_{\rm v}= 180$ meV \cite{Spano2010}). Shifts of this magnitude have been recently measured \cite{Schwartz2013}. 

\section{Conclusions}

In this work, we have developed a theoretical framework to describe the spectroscopy of organic microcavities based on the recently-introduced Holstein-Tavis-Cummings model \cite{Cwik2014,Spano2015,Herrera2016}, and a Lindblad description of the dissipative dynamics of organic polaritons. The model provides a microscopic interpretation for several observed features in the absorption and emission spectra of organic microcavities in the vicinity of the spectral region associated with bare molecular transitions. Several experimental features are given a consistent interpretation here for the first time. We do this by introducing the concept of {\it dark vibronic polaritons}, a new class of light-matter excitations that have no dipole oscillator strength, making them invisible in direct bound mode absorption \cite{Barnes1986,Matterson2001}, but which nevertheless have strong emission signatures in photoluminescence. We identify two types of dark vibronic polaritons for Rabi frequencies $\sqrt{N}\Omega\approx 2\,\omega_{\rm v}$, where $\omega_{\rm v}$ is the intramolecular vibration frequency. There is a set of non-degenerate $X$-type states that are dark in dipole absorption mostly due to destructive interference effects, and a set of $Y$-type dark vibronic polaritons, each set being are $(N-1)$-fold degenerate in the absence of energetic disorder. Dark vibronic polaritons of the $Y$ type are invisible in dipole absorption because they contain single-particle and two-particle material components that have no transition dipole moment with the absolute ground state of the system (no electronic, vibrational or cavity excitation). 

The introduced theoretical framework can be used to understand the photophysics of organic microcavities. In particular, we showed how the photoluminescence spectra can be divided into components associated with photon leakage transitions that leave the material with $\nu\geq 0$ vibrational excitations. Therefore, within the homogeneous model considered in this work,  emission near the lower polariton frequency can be due to direct {\it radiative} transitions (photon leakage) from polariton states at higher energies that leave the material with $\nu\geq 1$ vibrational quanta.  Non-radiative relaxation is thus not the only mechanism that can  account for  the observed emission enhancements at the lower polariton frequency, under off-resonant pumping \cite{Schwartz2013,Ballarini2014}. For large molecular ensembles in metallic cavities, we expect sub-picosecond radiative relaxation to be the dominant polariton decay channel, due to a predicted size-enhanced  fluorescence rate into bound modes of the nanostructure, and the short photon lifetime of low-$Q$ microcavities \cite{Hobson2002}.

 In the parameter regime where dark vibronic polaritons form near of the bare molecular transition ($\sqrt{N}\Omega\approx 2\,\omega_{\rm v}$),  material states that are commonly considered to be uncoupled from the cavity field (dark excitons) can borrow an appreciable photonic component from higher energy two-particle diabatic vibronic polaritons. In other words, there are no collective material states that remain uncoupled from the cavity field. Eventually, as the Rabi coupling increases far beyond the threshold for polaron decoupling \cite{Herrera2016}, the cavity spectra is expected to simplify. The study of the spectral changes in absorption and emission as the cavity transitions from the regime of dark vibronic polariton formation to the polaron decoupling regime is currently underway.

Possible experimental tests of the predicted photophysics of organic microcavities can include a comparison between  polariton photoluminescence into bound modes of the nanostructure versus the conventional leakage photoluminescence into free space. If the  external pump field is only slightly blue detuned relative to the lower polariton frequency, then dark vibronic polaritons are not populated and their effects on emission can be suppressed \cite{Herrera2016-PRL}. We find that the ratio between the vibrationless emission peak from (0-0) and the first vibronic sideband (0-1), associated with a transition from the lower polariton into the ground electronic manifold with one quantum of vibration, is different when measured via leakage photoluminescence in comparison with bound photoluminescence. Interestingly, only the bound mode 0-0/0-1 ratio is indicative of cavity-induced coherence length in the ensemble \cite{Spano2015}, resembling the behaviour of molecular aggregates in free space \cite{Spano2011}. 
 
In summary, we have demonstrated that a simple homogeneous treatment of the system with purely radiative relaxation can qualitatively describe several observed features of the spectroscopy of organic microcavities. More quantitative comparisons would require further extensions of the model, in order to account for long-range electrostatic coupling between emitters, strong energetic disorder, multi-mode cavity couplings, and more general radiative and non-radiative reservoirs. The model can be further extended to describe microcavities in the ultrastrong coupling regime \cite{Schwartz2011,Kena-Cohen2013,Gambino2015} having arbitrary mode dispersion.  Our work thus paves the way for the development of novel nonlinear optical devices \cite{Herrera2014,kowalewski2016cavity,Saurabh2016,Kowalewski2016}, chemical reactors \cite{Hutchison:2012,Herrera2016}, and optoelectronic devices \cite{Feist2015,Schachenmayer2015,Orgiu2015,yuen2016} that can be enhanced by quantum optics.  
\\

\acknowledgements
We thank Bill Barnes for providing  the photoluminescence data from Ref. \cite{Hobson2002},  Marina Litinskaya for technical discussions, and Jonathan Keeling for comments. F. Herrera is supported by CONICYT through  PAI N$^{\rm o}$ 79140030 and Fondecyt Iniciaci\'{o}n N$^{\rm o}$ 11140158. F. C. Spano is supported by the NSF, Grant N$^{\rm o}$ DMR-1505437.

\bibliographystyle{unsrt}
\bibliography{DVP}

\appendix
\section{Symmetry of the HTC model}
\label{app:symmetry}

Here we prove that 
\begin{equation}\label{eq:invariance}
{\rm e}^{-i\hat S}\hat H_{LM}{\rm e}^{i\hat S} = \hat H_{LM}.
\end{equation}
In order to do so, it is sufficient to show that the commutator $[\hat S,\hat H_{LM}]$ vanishes. Using the definitions in Eqs. (\ref{eq:HLM}) and (\ref{eq:S}),  
we can write this commutator as
\begin{eqnarray}\label{eq:commutator}
\lefteqn{[\hat S,\hat H_{LM}]=}\nonumber\\
&& \frac{1}{2}\sum_{\nu\nu'}\left[ (-1)^{-\nu}\Omega_{\nu\tilde \nu'}-(-1)^{\nu'}\Omega_{\tilde \nu \nu'}\right]\ket{e\,\tilde \nu\,0_c}\bra{e\,\tilde \nu'\,0_c}\nonumber\\
&&+\frac{1}{2}\sum_{\nu\nu'}\left[ (-1)^{-\nu'}\Omega_{\tilde \nu' \nu}-(-1)^{\nu}\Omega_{ \nu' \tilde \nu}\right]\ket{g\, \nu'\,1_c}\bra{g\,\nu\,1_c}.\nonumber\\
\end{eqnarray}
We use the following symmetry property of Frank-Condon factors under the exchange of vibrational indices \cite{Barnett-Radmore}
\begin{equation}\label{eq:FCF symmetry}
\braket{\nu'|\tilde \nu}  = (-1)^{\nu-\nu'}\braket{\nu|\tilde \nu'},
\end{equation}
to show that the square brackets in the first and second line of Eq. (\ref{eq:commutator}) can be rewritten to give
\begin{equation}\label{eq:first bracket}
 (-1)^{-\nu}\Omega_{\nu\tilde \nu'}-(-1)^{\nu'}\Omega_{\tilde \nu \nu'} = (-1)^{\nu}\left[\Omega_{\nu\tilde \nu'}-\Omega_{ \tilde \nu' \nu}\right]=0,
\end{equation}
and
\begin{equation}\label{eq:second bracket}
 (-1)^{-\nu'}\Omega_{\tilde \nu' \nu}-(-1)^{\nu}\Omega_{ \nu' \tilde \nu} = (-1)^{\nu'}\left[\Omega_{\tilde \nu'\nu}-\Omega_{\nu\tilde\nu'}\right] = 0,
\end{equation}
which proves the desired commutation property.

\section{Proof that $\bra{G}\hat \mu\ket{X} =0$}
\label{app:muX}

In this section we prove that a zero-energy vibronic polariton eigenstate of the HTC Hamiltonian $\hat{\mathcal{H}}$ [Eq. (\ref{eq:HTC})], in a frame rotating at the cavity frequency $\omega_c$, has no dipole oscillator strength for transitions from the absolute ground state of the cavity $\ket{G}\equiv \ket{g_1\,0_1,g_2\,0_2,\ldots,g_N\,0_N}\ket{0_c}$.

We write the transition dipole operator as $\hat \mu = \hat \mu^{(+)}+\hat \mu^{(-)}$, where $\hat \mu^{(+)} = \mu \sum_n\ket{g_n}\bra{e_n}$ and $\hat \mu^{(-)}=[\hat \mu^{(+)}]^\dagger$, with $\mu$ being the single-particle transition dipole moment. Denoting as $\ket{X}$ the vibronic polariton eigenstate with zero eigenvalue in the cavity frame, we have
\begin{equation}\label{eq:HX=0}
\hat{\mathcal{H}}\ket{X}=0.
\end{equation}
Left-multiplying this expression by the state $\bra{G}\hat a$ gives
\begin{eqnarray}
\bra{g_1\,0_1,g_2\,0_2,\ldots,g_N\,0_N}\bra{1_c}\sum_n\left\{\ket{g_n}\bra{e_n}\hat a^\dagger\right.\nonumber\\
\left. +\ket{e_n}\bra{g_n}\hat a\right\}\ket{X}=0.
\end{eqnarray}
Only the first term in the curly bracket contributes to the left-hand side of this equality, giving
\begin{equation}\label{eq:mu plus}
\bra{G}\hat \mu^{(+)}\ket{X}=0.
\end{equation}
This expression gives the desired result $\bra{G}\hat \mu\ket{X} = 0$, by noting that $\bra{G}\hat \mu^{(-)}=0$.

\section{Input-Output Relations}
\label{app:input output}

We review here the input-output theory for transmission and reflection spectra of a two-sided planar cavity. We follow the so-called quasi-mode approximation, in which orthogonal degrees of freedom inside and outside the cavity are defined, these being weakly coupled to each other via finite transmission through the cavity mirrors \cite{Carmichael-book1,Savona1999}.  The formalism is an alternative to the field quantization approach based on the classical Green's function of the problem \cite{Knoll2001}. We follow the Schrodinger picture approach from Carmichael {\it et al.} \cite{Carmichael-book1}, which differs the Heisenberg picture approach from Collet and Gardiner \cite{Gardiner1985,Ciuti2006} in that no time-reversed solutions are invoked. 

The dynamics of an empty cavity coupled to the electromagnetic reservoirs at each external side of the cavity mirrors is given by the Hamiltonian $\hat H = \hat H_S +\hat H_{R}+\hat H_{SR}$,
 where 
 \begin{eqnarray}
 \hat H_S &=& \sum_{\bbeta}\omega_{\bbeta}\,\hat a_{\bbeta}^\dagger \hat a_{\bbeta}\label{eq:HS}\nonumber\\
 \hat H_{R}&=&  \sum_{i\bbeta}\int_0^\infty d\omega \,\omega \left[\hat l_{i\bbeta}^\dagger(\omega) \hat l_{i\bbeta}(\omega)+\hat r_{i\bbeta}^\dagger(\omega) \hat r_{i\bbeta}(\omega)\right]\label{eq:HR}\nonumber\\
 \hat H_{SR}&=&\sum_{i\bbeta} \int_0^\infty d\omega \,\kappa_{i}^*(\omega) \,\hat a_{\bbeta}\left[ \hat l^\dagger_{i\bbeta}(\omega)+\hat r^\dagger_{i\bbeta}(\omega)\right]+{\rm H.c.}\label{eq:HSR}\nonumber\\
 \end{eqnarray}
where the operators $\hat l_{i\bbeta}(\omega)$ and $\hat r_{i\bbeta}(\omega)$ are continuous reservoir field operators that annihilate a free-space photon with frequency $\omega$ and in-plane wave vector $\bbeta$ propagating to the left and to the right relative to the cavity axis, on regions $i =1$ (left-side of cavity) and region $i = 2$ (right-side). 
We implicitly assumed that all fields involved have TE polarization $\hat \e_{\beta}$ and only the lowest TE cavity mode is taken into account. We have also used the fact that transmission through the cavity mirror preserves the in-plane wave vector $\bbeta$ of the field away from the mirror edges. Since the cavity length $L$ is sub-wavelength, higher order TE modes with are far detuned from the electronic transition frequencies of interest. The system-reservoir couplings $\kappa_{i}(\omega)$ determine the cavity decay rate at each mirror, as we show below. The reservoir operators satisfy the equal-time commutation relations of the form
\begin{eqnarray}
\left[\hat l_{i\bbeta}(\omega),\hat l^\dagger_{k\bbeta'}(\omega')\right] =\left[\hat r_{i\bbeta}(\omega),\hat r^\dagger_{k\bbeta'}(\omega')\right]&=& \delta_{ik}\delta(\omega-\omega')\nonumber\\
\left[\hat l_{i\bbeta},\hat r^\dagger_{k\bbeta'}\right]=\left[\hat a_{\bbeta},\hat r^\dagger_{k\bbeta'}\right]=\left[\hat a_{\bbeta},\hat l^\dagger_{k\bbeta'}\right]&=&0.
\end{eqnarray}
Cavity and reservoir operators evaluated at different times do not necessarily commute.

We can avoid referring to the electric field normalization outside the cavity by defining the electric field operator in units of photon flux for $i$-th region  $\hat{\mathcal{E}}_i^{(+)}(\x,t) = \sum_{\bbeta} \, {\rm e}^{i\bbeta \cdot \x_\parallel}\hat{\mathcal{E}}_{i\bbeta}(z,t)$, with
\begin{eqnarray}\label{eq:external fields}
\sqrt{2\pi}\,\hat{\mathcal{E}}^{(+)}_{i\bbeta}(z,t) &=&\int_0^\infty d\omega
\, \hat l_{i\bbeta}(\omega,t){\rm e}^{-i\omega\zeta_i(z)/c}\nonumber\\
&&+\int_0^\infty d\omega\,\hat r_{i\bbeta}(\omega,t){\rm e}^{i\omega\zeta_i(z)/c},
\end{eqnarray}
where $\zeta_1(z) \equiv z\cos\theta_1<0$ is the projected wavefront distance relative to the z axis (cavity axis) with $\theta_1$ being the incidence angle in region 1. For region 2 we have the scaled position $\zeta_2(z) \equiv (z-L)\cos\theta_2>0$, where $\theta_2$ is the incidence angle relative to the cavity axis. In what follows, we evaluate the frequency integrals by writing $\omega = \omega_{\bbeta} +\Delta$, where $\Delta$ is the detuning from the cavity resonance frequency $\omega_{\bbeta}$. The resulting expressions are simplified when written in terms of rotating frame operators defined as $\tilde A(t)\equiv \hat{A}(t){\rm e}^{i\omega_{\bbeta}t}$.

The Heisenberg equations of motion for the rotating-frame operators $\tilde l_{i\bbeta}(\Delta,t)$ and $\tilde r_{i\bbeta}(\Delta,t)$ in each region can be formally solved to read
\begin{eqnarray}\label{eq:reservoir operators}
\tilde l_{i\bbeta}(t) & =& \tilde l_{i\bbeta}(0){\rm e}^{-i\Delta t}-i{\kappa_{i}^*(\Delta)}\int_0^t dt'\;{\rm e}^{-i\Delta(t-t')}\tilde a_{\bbeta}(t')\nonumber\\
\tilde r_{i\bbeta}(t) & =& \tilde r_{i\bbeta}(0){\rm e}^{-i\Delta t}-i{\kappa_{i}^*(\Delta)}\int_0^t dt'\;{\rm e}^{-i\Delta(t-t')}\tilde a_{\bbeta}(t'),\nonumber\\
\end{eqnarray}
where we have omitted for simplicity the frequency dependence of the reservoir operators.
These formal solutions are then inserted into the rotating-frame version of Eq. (\ref{eq:external fields}) to obtain a field of the form 
\begin{equation}\label{eq:external}
\hat{\mathcal{E}}^{(+)}_{i\bbeta} = \hat{\mathcal{E}}^{(+)}_{i\bbeta L}+\hat{\mathcal{E}}^{(+)}_{i\bbeta R}+\hat{\mathcal{E}}^{(+)}_{i\bbeta S},
\end{equation}
 where 
\begin{equation}\label{eq:free left}
\hat{\mathcal{E}}^{(+)}_{i\bbeta L}(z,t) = {\rm e}^{-i\omega_{\bbeta}(t+\zeta_i/c)}\int \frac{d\Delta}{\sqrt{2\pi}}\;\tilde l_{i\bbeta}(\Delta,0){\rm e}^{-i\Delta(t+\zeta_i/c)},
\end{equation}
and 
\begin{equation}\label{eq:free right}
\hat{\mathcal{E}}^{(+)}_{i\bbeta R}(z,t) = {\rm e}^{-i\omega_{\bbeta}(t-\zeta_i/c)}\int \frac{d\Delta}{\sqrt{2\pi}}\;\tilde r_{i\bbeta}(\Delta,0){\rm e}^{-i\Delta(t-\zeta_i/c)}
\end{equation}
are free fields propagating to the left and to the right along the $z$ axis, respectively. 
As expected for free fields, these expressions give $[\hat{\mathcal{E}}^{(+)}_{i\bbeta L}(z,t),\hat{\mathcal{E}}^{(-)}_{i\bbeta L}(z,t')]=[\hat{\mathcal{E}}^{(+)}_{i\bbeta R}(z,t),\hat{\mathcal{E}}^{(-)}_{i\bbeta R}(z,t')]=\delta(t-t')$. The reservoir operators $\tilde l_{i\bbeta}(\Delta)$ and $\tilde r_{i\bbeta}(\Delta)$ evaluated at $t =0$,  become input field operators in the formalism by Collet and Gardiner \cite{Gardiner1985}.  The total field $\hat{\mathcal{E}}^{(+)}_{i\bbeta}$ also has a source contribution that depends on the intracavity field and can be written as
\begin{eqnarray}\label{eq:source field}
\lefteqn{\sqrt{2\pi}\,\hat{\mathcal{E}}^{(+)}_{i\bbeta S}(z,t) =}\\
&&-i{\rm e}^{-i\omega_{\bbeta}(t+\zeta_i/c)} \int d\Delta\;{\kappa_i^*(\Delta)}\int_0^t dt'\;{\rm e}^{-i\Delta(t-t'+\zeta_i/c)}\tilde a_{\bbeta}(t')\nonumber\\
&&-i\,{\rm e}^{-i\omega_{\bbeta}(t-\zeta_i/c)} \int d\Delta\;{\kappa_i^*(\Delta)}\int_0^t dt'\;{\rm e}^{-i\Delta(t-t'-\zeta_i/c)}\tilde a_{\bbeta}(t').\nonumber
\end{eqnarray}
Causality ensures that either on the left side or right side of the cavity, only one term in Eq. (\ref{eq:source field}) contributes to the source field. Physically, it is not possible for a right-propagating field far to the left cavity mirror ($\zeta_1\rightarrow -\infty$), for example, to depend on the intracavity field operator $\tilde a_{\bbeta}$ because it has yet to reach the mirror. Therefore its contribution to the source field $\hat{\mathcal{E}}^{(+)}_{i\bbeta S}$ must vanish. The left-propagating contribution at the same distant location can however depend on the intracavity field after reflection at the left cavity mirror. Such causal behavior is obtained in a simple form by assuming a frequency-independent coupling of the form $\kappa_i(\omega) \approx \kappa_{i}(\omega_{\bbeta})$ over the frequency range $\omega_{\bbeta}-\delta_c<\omega<\omega_{\bbeta}+\delta_c$, where $\delta_c$ is a cut-off frequency, which gives  $\kappa_i(\Delta)=\kappa_i(0)$ in the rotating frame of the cavity mode. The frequency integrals in Eq. (\ref{eq:source field}) can be easily evaluated by setting $\delta_c\rightarrow \infty$. The source field in region 1 ($\zeta_1<0$) thus reads
\begin{eqnarray}
\lefteqn{\hat{\mathcal{E}}^{(+)}_{1\bbeta S}(z,t)=} \\
&&-i\sqrt{2\pi}\;\kappa_1^*(0) \left[{\rm e}^{-i\omega_{\bbeta}(t+\zeta_1/c)}\int_0^t dt' \tilde a_{\bbeta}(t')\delta(t+\zeta_1/c-t')\right.\nonumber\\
&&\left.+\; {\rm e}^{-i\omega_{\bbeta}(t-\zeta_1/c)}  \int_0^t dt' \tilde a_{\bbeta}(t')\delta(t-\zeta_1/c-t')\right].\nonumber
\end{eqnarray}
The second time integration (right-propagating wave) vanishes because the peak of the delta function lies outside the integration range $(0,t)$ for $\zeta_1\leq0$. Only the left-propagating contribution survives, giving the source field
\begin{equation}\label{eq:source 1}
\hat{\mathcal{E}}^{(+)}_{1\bbeta S}(z,t) = -i{\rm e}^{-i\omega_{\bbeta}(t+\zeta_1/c)} \sqrt{2\pi}\;\kappa_1^*(0)\,\tilde a_{\bbeta}(t-|\zeta_1|/c),
\end{equation}
which depends on the intracavity field evaluated at the retarded time $t-|\zeta_1|/c$, as expected from causality. We made the change of variable $\tau=t+\zeta_1/c-t'$ in evaluating the time integral. The source contribution in region 2 ($\zeta_2\geq 0$) can be obtained in an analogous way to give
\begin{equation}\label{eq:source 2}
\hat{\mathcal{E}}^{(+)}_{2\bbeta S}(z,t) = -i{\rm e}^{-i\omega_{\bbeta}(t-\zeta_2/c)} \sqrt{2\pi}\;\kappa_2^*(0)\,\tilde a_{\bbeta}(t-\zeta_2/c).
\end{equation}

The coupling constants $\kappa_i(0)$ evaluated at the cavity resonance frequency $\omega_{\bbeta}$ are related to the damping rates $\gamma_i$ that determine the cavity linewidth \cite{Carmichael-book1}. Starting from the Heisenberg equation of motion for the slowly-varying amplitude
\begin{equation}
\frac{d}{dt}\tilde a_{\bbeta} = -i\sum_i\int_{-\infty}^\infty d\Delta\, \kappa_i(\Delta)\left[\hat l_{i\bbeta}(\Delta,t)+\hat r_{i\bbeta}(\Delta,t)\right],
\end{equation}
after inserting the formal solutions from Eq. (\ref{eq:reservoir operators}), we arrive at the Langevin equation
\begin{eqnarray}\label{eq:Langevin}
\frac{d}{dt}\tilde a_{\bbeta} &=& -\int_0^t dt'\;\left[K_1(t-t') +K_2(t-t')\right]\tilde a_{\bbeta}(t')\nonumber\\
&&+\hat L_{1\beta}(t)+\hat R_{1\beta}(t)+\hat L_{2\beta}(t)+\hat R_{1\beta}(t),
\end{eqnarray}
with  dissipation kernel 
\begin{eqnarray}\label{eq:dissipation kernel}
K_i(t-t') &=& \int_{-\infty}^\infty \,d\Delta \,|\kappa_i(\Delta)|^2{\rm e}^{-i\Delta(t-t')},
\end{eqnarray}
and source operators 
\begin{eqnarray}
\hat L_{i\bbeta}(t) &=& -i\int_{-\infty}^\infty d\Delta \,\kappa_i(\Delta)\tilde l_{i\bbeta}(\Delta,0){\rm e}^{-i\Delta t}\\
\hat R_{i\bbeta}(t) &=& -i\int_{-\infty}^\infty d\Delta \, \kappa_i(\Delta)\tilde r_{i\bbeta}(\Delta,0){\rm e}^{-i\Delta t}.
\end{eqnarray}
These are determined by the input reservoir operators $\tilde l_{i\bbeta}(\Delta)$ and $\tilde r_{i\bbeta}(\Delta)$ at each side of the cavity. In the Markovian approximation for the cavity-mirror coupling, the dissipation kernels become $K_i(t-t') = (2\pi)|\kappa_i(0)|^2\delta(t-t')$. Setting
\begin{equation}
\gamma_i = 2\pi |\kappa_i(0)|^2, 
\end{equation}
and using the relation  $\int_0^td\tau f(\tau)\delta(\tau)=f(0)/2$ to integrate the time kernels, gives a time-local equation of motion 
\begin{eqnarray}\label{eq:Langevin equation}
\frac{d}{dt}\tilde a_{\bbeta}(t) &=& -(\gamma/2)\,\tilde a_{\bbeta}(t)+\sqrt{\gamma_1}\,\tilde F_{1\bbeta}(t)+\sqrt{\gamma_2}\,\tilde F_{2\bbeta}(t)\nonumber\\
\end{eqnarray}
where $\gamma = (\gamma_1+\gamma_2)$ is the cavity decay rate, and the Langevin noise operators are given by
\begin{eqnarray}\label{eq:Langevin input}
\tilde F_{i\beta}(t) &=& -\frac{i}{\sqrt{2\pi}}{\rm e}^{i\phi_i}\int d\Delta \;\left[\tilde l_{i\bbeta}(\Delta,0)+\tilde r_{i\bbeta}(\Delta,0)\right]{\rm e}^{-i\Delta t},\nonumber\\
\end{eqnarray}
where we have defined the phase $\phi_i$ as $\kappa_i(0)=|\kappa_i(0)|{\rm e}^{i\phi_i}$. It is well-known that the Langevin equation in Eq. (\ref{eq:Langevin equation}) can be rewritten for coherent input fields as the Lindblad quantum master equation  \cite{Carmichael-book1} 
\begin{equation}\label{eq:photon Lindblad}
\dot{ \hat \rho} = -i\left[\hat H_S,\hat \rho\right] + \gamma\left(\hat a_{\bbeta} \hat \rho \hat a^\dagger_{\bbeta} -\frac{1}{2}\{\hat a_{\bbeta}^\dagger\hat a_{\bbeta},\hat \rho \}\right),
\end{equation}
where $\hat \rho$ is intracavity photon reduced density matrix and the curly brackets denote an anti-commutator. The Langevin noise operators $F_{i\bbeta}(t)$ are represented by a near resonant driving Hamiltonian term, such that the intracavity Hamiltonian reads
\begin{eqnarray}
\hat H_S =\omega_{\bbeta}\hat a_{\bbeta}^\dagger\hat a_{\bbeta}+\mathcal{E}_1^*\hat a_{\bbeta}{\rm e}^{-i\Delta_{\bbeta} t}+\mathcal{E}_1\hat a_{\bbeta}^\dagger{\rm e}^{i\Delta_{\bbeta} t}
\end{eqnarray}
where $\mathcal{E}_1 = \sqrt{\gamma_1}\beta_1$ is the driving strength corresponding to an input coherent state with complex amplitude $\beta_1$ on mirror 1, detuned from the cavity resonance by $\Delta_{\bbeta}$, assuming driving on region 1 only.

Transmission and reflection experiments measure the spectrum of the external electric field $\hat{\mathcal{E}}_i^{(+)}(\x,t)$ for a given in-plane momentum $\bbeta$ at a fixed position $\x=\x_i$.  For driving on region 1, the reflection spectra $R(\omega)$ is determined by the Fourier transform of the normalized first-order correlation function \cite{Walls2008}
\begin{equation}\label{eq:g1}
g_1^{(1)}(t,\tau) =  \frac{\langle \hat{\mathcal{E}}_{1\bbeta}^{(-)}(t+\tau)\hat{\mathcal{E}}_{1\bbeta}^{(+)}(t)\rangle}{\langle \hat{\mathcal{E}}_{1\bbeta}^{(-)}(t)\hat{\mathcal{E}}_{1\bbeta}^{(+)}(t)\rangle}.
\end{equation}
The transmission spectra $T(\omega)$ would be proportional to the Fourier transform of the field correlation function $\langle \hat{\mathcal{E}}_{2\bbeta}^{(-)}(t+\tau)\hat{\mathcal{E}}_{2\bbeta}^{(-)}(t)\rangle$, corresponding to region 2.  

The field expansion introduced in Eq. (\ref{eq:external}) shows that in general the output spectrum (either $R(\omega)$ of $T(\omega)$) depends on the correlations between the reservoir and intracavity field operators. However,  these can be neglected at optical frequencies for thermal reservoirs \cite{Carmichael-book1}, giving the stationary correlation function ($\tau>0$)
\begin{eqnarray}
g_i^{(1)}(t,\tau)  &\approx& \frac{\left[\langle \hat{\mathcal{E}}_{i\bbeta L}^{(-)}(t+\tau)\hat{\mathcal{E}}_{i\bbeta L}^{(+)}(t)\rangle\right.+\langle \hat{\mathcal{E}}_{i\bbeta R}^{(-)}(t+\tau)\hat{\mathcal{E}}_{i\bbeta R}^{(+)}(t)\rangle\left.\right]}{\langle \hat{\mathcal{E}}_{i\bbeta}^{(-)}(t)\hat{\mathcal{E}}_{i\bbeta}^{(+)}(t)\rangle}\nonumber\\
&&+\frac{\pi\gamma_i\;\langle \tilde{a}_{\bbeta}^\dagger(t+\tau)\tilde{a}_{\bbeta}(t)\rangle}{\langle \hat{\mathcal{E}}_{i\bbeta}^{(-)}(t)\hat{\mathcal{E}}_{i\bbeta}^{(+)}(t)\rangle}
\end{eqnarray}

The steady state spectrum is obtained by taking the limit $t\rightarrow \infty$ in Eq. (\ref{eq:g1}). Therefore, once the input field spectrum is known (driving laser), the stationary transmission and reflection spectra are completely determined by the intracavity two-time correlation function $\langle \tilde{a}_{\bbeta}^\dagger(t+\tau)\tilde{a}_{\bbeta}(t)\rangle$, which can be evaluated from the Lindblad master equation or Langevin equation for the system evolution, together with the quantum regression theorem \cite{Carmichael-book1}. 

\subsection{Transmission and Reflection}

Before discussing light absorption by a molecular medium embedded in the cavity, let us consider a two-mirror cavity driven by a stationary coherent input field with (complex) amplitude $\langle \hat F_{1\bbeta}\rangle \neq 0$ on region 1, i.e., $\langle \hat F_{2\bbeta}\rangle = 0$ in Eq. (\ref{eq:Langevin equation}). The steady state amplitude of the intracavity field is thus
\begin{equation} \label{eq:a ss}
\langle a_{\bbeta} \rangle_{\rm ss} = -\frac{2\sqrt{\gamma_1}}{\gamma}\langle\hat F_{1\bbeta}\rangle. 
\end{equation}
From Eqs. (\ref{eq:free left}), (\ref{eq:free right}), (\ref{eq:source 1}) and (\ref{eq:Langevin input}), we can write the electric field flux amplitude at the reflecting mirror ($\zeta_1=0$) as
\begin{equation}\label{eq:reflection}
|\langle \hat{\mathcal{E}}_{1\bbeta}^{(+)}\rangle| = |\langle \hat F_{1\bbeta}\rangle| \left(1-\frac{2\gamma_1}{\gamma}\right)=|\langle \hat F_{1\bbeta}\rangle|\left(\frac{\gamma_2-\gamma_1}{\gamma_2+\gamma_1}\right).
\end{equation}
A similar calculation for the electric field amplitude at the transmitting mirror ($\zeta_2=0$) gives
\begin{equation}\label{eq:transmission}
|\langle \hat{\mathcal{E}}_{2\bbeta}^{(+)}\rangle| = |\langle \hat F_{1\bbeta}\rangle| \,\frac{\sqrt{\gamma_1\gamma_2}}{\gamma_1+\gamma_2}.
\end{equation}
These expressions simply state the fact that for identical mirrors ($\gamma_1=\gamma_2$), the reflected field vanishes and the cavity acts as a perfect transmission filter, which is known from classical interferometry \cite{Macleod2013}. Since there is no absorption in the cavity, conservation of photon flux can be stated as
\begin{equation}\label{eq:R+T=1}
|\langle\hat{\mathcal{E}}_{1\bbeta}^{(+)}\rangle|^2+|\langle\hat{\mathcal{E}}_{2\bbeta}^{(+)}\rangle|^2 = |\langle \hat F_{1\bbeta}\rangle|^2.
\end{equation}
Normalizing the reflected and transmitted fluxes by the input photon flux, results in the identity \[R+T=1,\] where $R$ and $T$ are the reflectivity and transmittivity of the cavity at a given input frequency. 

\subsection{Cavity Absorption}
\label{app:absorption}

Let us now consider a molecular medium embedded in a two-side cavity. The interaction between the molecular transition dipole moment operator of the $n$-th molecule in the ensemble $\hat \mu_n$ and the local electric field operator $\hat{\mathcal{E}}(\x_n)$ is represented, ignoring a global phase, by the interaction term
\begin{equation}\label{eq:mu dot E}
\hat H_I = \sum_n \hat \mu_n \,\hat{\mathcal{E}}(\x_n) =  \sum_n \hat \mu_n^{(-)} \,\hat{\mathcal{E}}^{(+)}(\x_n)+{\rm H.c.},
\end{equation}
where in the second equality we use $\hat \mu  = \hat \mu^{(+)}+\hat \mu^{(-)}$ and adopted the rotating wave approximation (RWA) \cite{Walls2008}. The local electric field at the molecular position $\x_n$ is given by superposition of multiple field modes of the microcavity. The field modes whose in-plane wave vector $\bbeta$ does not exceed the cutoff for total internal reflection are able to leak out of the nanostructure through the mirrors. These so-called {\it radiative} modes are involved in transmission and reflection measurements \cite{Hobson2002}. The dynamics of the radiative modes is determined by the field operator $\hat a_{\bbeta}$, which satisfies Eq. (\ref{eq:Langevin equation}) for an empty cavity. 

For those field modes whose in-plane wave vector exceed the cutoff for total internal reflection, photons are unable to leak through the mirrors and remain bound to the intracavity space. Bound modes are mostly confined to the bulk of the dielectric medium or at the mirror-dielectric interface, and represent a decay channel for molecular electronic  transitions 
that can reduce the efficiency of light-emitting devices \cite{Matterson2001}. The degree in which a bound mode is either bulk or a surface mode depends on the field polarization and the distance between the mirrors \cite{Torma2015}. We use the operator $\hat b_{\bbeta'}(\omega)$  to represent a bound mode with in-plane wavevector $\bbeta'$, and write the local intracavity (positive frequency) electric field as the superposition
\[\hat{\mathcal{E}}^{(+)}(\x_n)= \hat{\mathcal{E}}_a^{(+)}(\x_n)+\hat{\mathcal{E}}_b^{(+)}(\x_n),\]
 where $\hat{\mathcal{E}}_a^{(+)}(\x_n)$ is the contribution to the electric field from radiative modes, and $\hat{\mathcal{E}}_b^{(+)}(\x_n)$ the contribution from bound modes. We treat radiative and bound modes differently in our model. We include only the radiative modes as part of the Holstein-Tavis-Cummings (HTC) Hamiltonian, which reads
 \begin{equation}\label{eq:HTC cavity}
 \hat{\mathcal{H}} = \hat H_{\rm C} + \hat H_{\rm M} + \hat V_a,
 \end{equation}
where $\hat H_{\rm C}$ is the free cavity Hamiltonian from Eq. (\ref{eq:HS}), $\hat H_{\rm M}$ is the Holstein  Hamiltonian involving electronic and vibrational degrees of freedom of the material, and $\hat V_{a} =\sum_n [\hat \mu_n^{(-)}\hat{\mathcal{E}}_a^{(+)}(\x_n)+{\rm H.c.}]$ is the cavity-matter coupling. The bound modes are considered as a zero-temperature electromagnetic reservoir that induces dissipation of the molecular dipoles via spontaneous emission. Expanding the bound electric field as 
\begin{equation}
\sqrt{2\pi}\,\hat{\mathcal{E}}_b^{(+)}(\x)= \sum_{\bbeta'}\int_0^\infty d\omega\, \hat b_{\bbeta'}(\omega)\,{\rm e}^{i\bbeta'\cdot \x},
\end{equation}
gives an effective system-reservoir coupling Hamiltonian of the form
\begin{equation}
\hat H_{\rm SR} = \sum_n\sum_{\bbeta'}\int_0^\infty\, d\omega\, g(\omega)\, \hat \sigma_n^-\,\hat b_{\bbeta'}(\omega)\,{\rm e}^{i\bbeta' \cdot \x_n}+{\rm H.c.},
\end{equation}
where $ \hat \sigma_n^{-} = \hat \mu_n^{(+)}/\mu$ is the electronic transition operator of the $n$-th molecule, $\mu$ is the transition dipole moment, and $g(\omega)$ is a coupling function. It is well-known \cite{Carmichael-book1,Walls2008,Breuer-book} that this system-reservoir interaction in the Born-Markov approximation gives a Lindblad quantum master equation for the system density matrix $\hat \rho$ of the form
\begin{equation}\label{eq:Lindblad general}
\frac{d}{dt}\hat \rho(t) = -i[\hat{\mathcal{H}},\hat \rho(t)]+ \mathcal{L}_a\left[\hat\rho (t)\right]+ \mathcal{L}_\mu[\hat \rho(t)],
\end{equation}
where $\mathcal{H}$ is the HTC Hamiltonian in Eq. (\ref{eq:HTC cavity}), $\mathcal{L}_a\left[\hat\rho (t)\right]$ is the cavity photon Lindblad dissipator given by the second term in Eq. (\ref{eq:photon Lindblad}), and $ \mathcal{L}_\mu[\hat \rho(t)]$ is the dissipator associated with dipole radiative decay into bound modes, which can be written as
\begin{equation}\label{eq:Lindblad dipole}
 \mathcal{L}_\mu[\hat \rho(t)] = \sum_n \gamma_n \left(\hat \sigma_n^{-}\hat \rho(t)\hat \sigma_n^{+}-\frac{1}{2}\{\hat \sigma_n^{+}\hat \sigma_n^{-},\hat \rho(t)\}\right),
\end{equation}
where $\gamma_n$ is single-emitter decay rate $\gamma_n$.

We can follow a derivation to the one described above for an empty cavity and formally solve the Heisenberg equation of motion for the slowly-varying variable $\tilde b_{\bbeta'}(\Delta,t)$, where $\Delta$ is a small detuning of the bound mode frequency $\omega_{\bbeta'}$ from the empty cavity frequency $\omega_{\bbeta}$. We use this formal solution to write the bound electric field amplitude as
\begin{eqnarray}\label{eq:Eb free source}
\lefteqn{\hat{\mathcal{E}}_b^{(+)}(\x,t)=-i\sum_{n,\bbeta'}{\rm e}^{i[\bbeta'\cdot (\x-\x_n)-\omega_{\bbeta'}t]}}\\
&&\times\int\,\frac{d\Delta}{\sqrt{2\pi}} g^*(\Delta)\int_0^t\,dt'\,{\rm e}^{-i\Delta(t-t')} \,\tilde \sigma_n^{-}(t').\nonumber
\end{eqnarray}
We neglect the free field contribution to $\hat{\mathcal{E}}_b^{(+)}$ because its mean amplitude vanishes for a zero-temperature electromagnetic reservoir, i.e.,  $\langle \tilde b_{\bbeta'}(\Delta,0)\rangle=0$, and its contribution to the field second moments is separable \cite{Carmichael-book1}. The bound field is thus determined by a sum over dipole sources, and its dynamics is given by the evolution of the slowly-varying transition operator $\tilde J_n^{(+)}(t)$. Assuming $g(\Delta)\approx g(0)$ and that $|\x-\x_n|\approx |\x|$ in the far field, we obtain
\begin{equation}
\hat{\mathcal{E}}_b^{(+)}(\x,t) = -i\sqrt{2\pi} \,g^*(0)\sum_{\bbeta'}\sum_n \tilde \sigma_n^{-}(t-|\x|/c)\,{\rm e}^{-i(\omega_{\bbeta'} t-\bbeta'\cdot \x)}.
\end{equation}
The fluorescence spectrum associated with spontaneously emitted light into bound modes by the molecules in the ensemble is thus determined by the Fourier transform of the first-order correlation function
\begin{equation}\label{eq:g1 bound}
g_b^{(1)}(t,\tau)=\frac{\langle \hat{\mathcal{E}}_b^{(-)}(\x,t+\tau)\hat{\mathcal{E}}_b^{(+)}(\x,t) \rangle}{\langle \hat{\mathcal{E}}_b^{(-)}(\x,t)\hat{\mathcal{E}}_b^{(+)}(\x,t) \rangle},
\end{equation}
  which can be evaluated from the evolution of the dipole coherence $\sum_n\langle \sigma_n^-(t)\rangle$, together with the quantum regression theorem \cite{Carmichael-book1}.  For simplicity, we assume that two-point dipole correlations of the form $\langle \sigma_n^{+}(\tau)\sigma_{m}^-(0)\rangle $ are negligibly small for the timescales of interest, which is consistent with the local decay approximation assumed in Eq. (\ref{eq:Lindblad dipole}).
  
We finally relate the adopted cavity QED approach to cavity emission \cite{Carmichael-book1} with the usual definition of  cavity absorption $A(\omega)$ in semiconductor microcavities. The microcavity is assumed to be driven with a continuous wave coherent field of frequency $\omega_p$, on the external side of one of the mirrors (Region 1). For simplicity of notation, we assume driving at normal incidence ($\bbeta = 0$). The photon flux (in Hz) of the driving field given by $|\langle \hat F_{1}\rangle|^2 $. In the absence of dipole decay, radiative or non-radiative, the reflected and transmitted photon fluxes, $|\langle \hat{\mathcal{E}}_1^{(+)}\rangle|^2$ and $|\langle \hat{\mathcal{E}}_2^{(+)}\rangle|^2$, respectively, should satisfy Eq. (\ref{eq:R+T=1}) for any $\omega_p$, which gives $R(\omega_p)+T(\omega_p)=1$ for the normalized reflection and transmission spectra. The coherent cavity-matter coupling $\hat V_a$ in Eq. (\ref{eq:HTC cavity}) does not alter this relation, but simply leads to a Rabi splitting in the strong coupling regime.  In a more realistic scenario, where molecular dipoles decay radiatively into bound modes of the microcavity structure, conservation of photon flux can be stated as 
\begin{equation}\label{eq:total flux}
|\langle \hat{\mathcal{E}}_1^{(+)}\rangle |^2+|\langle \hat{\mathcal{E}}_2^{(+)}\rangle |^2+|\langle \hat{\mathcal{E}}_b^{(+)}\rangle|^2 = |\langle \hat F_{1}\rangle|^2,
\end{equation}
 for any given $\omega_p$. We have implicitly assumed that all bound mode fluorescence is collected in the far field. Normalizing the above equality by the input flux gives the usual definition the cavity absorption spectra
 \begin{equation}
 A = 1-R-T.
 \end{equation}
We thus show that the absorbed photon flux $A(\omega_p)$ at the driving frequency $\omega_p$ is determined by the normalized bound fluorescence intensity. In physical terms, fluorescence into bound modes of the microcavity attenuates the transmission and reflection of a driving laser. More generally, any process that leads to dipole decay, such as non-radiative molecular relaxation, would contribute to such field attenuation and contribute to the absorption spectra. However, we expect radiative relaxation to occur at a rate on the order of $\sqrt{N}\gamma_{\rm e}$ due to the collective character of the electronic degrees of freedom in the strong cavity-matter coupling regime. Such size-enhancements of the bound fluorescence rate may reduce the contribution of non-radiative processes to absorption. 


\section{Spectrum of Polariton Fluctuations}
\label{app:fluctuations}

We derive here a general formula to compute two-time autocorrelations of the form \[\langle \hat O_1(t+\tau)\hat O_2(t)\rangle,\] as needed in the computation of the leakage photoluminescence (LPL) of weakly  driven cavities. The starting point of the derivation is the quantum regression formula \cite{Carmichael-book1,Breuer-book} 
\begin{equation}\label{eq:regression formula}
\langle\hat O_1(t+\tau)\hat O_2(t)\rangle = {\rm Tr}_S\left\{\hat O_1(0){\rm e}^{\mathcal{L}\tau}\left[\hat O_2(0)\hat \rho(t)\right]  \right\},
\end{equation}
where $\hat \rho(t)$ is the reduced density matrix of the system (organic cavity) that evolves according to  $\hat \rho(t) ={\rm e}^{\mathcal{L}t}\hat \rho(0)$, where $\hat \rho(0)$ is the initial state of the system and $\mathcal{L}$ is the Liouville super-operator \cite{Carmichael-book1,Breuer-book}. Using regression formula [Eq. (\ref{eq:regression formula})] thus involves solving the master equation twice; first with respect to the absolute time $t$, starting from the initial condition $\hat \rho(0)$ in order to obtain the evolved state $\hat \rho(t)$, and then solve 
\begin{equation}\label{eq:sigma equation}
\frac{d}{d\tau}\hat \sigma_t(\tau) = \mathcal{L}\,\hat \sigma_t(\tau),
\end{equation}
with respect to the delay time $\tau$. This $\tau$-dependent equation has the formal solution 
\begin{equation}\label{eq:sigma t}
\hat \sigma_t(\tau) = {\rm e}^{\mathcal{L}\tau}\hat \sigma_t(0),
\end{equation}
for the $t$-dependent initial state $\hat \sigma_t(0)= \hat O_2\hat \rho(t)$. 
The required two-time correlation function in Eq. (\ref{eq:regression formula}) thus be rewritten as
\begin{equation}\label{eq:TPC sigma}
\langle\hat O_1(t+\tau)\hat O_2(t)\rangle = {\rm Tr}_S\left\{\hat O_1(0)\hat \sigma_t(\tau)  \right\}.
\end{equation}

\subsection{No-Quantum-Jump (NQJ) Approximation}

Assuming a standard Lindblad form for the Liovuille super-operator $\mathcal{L}$ \cite{Breuer-book}
\begin{equation}
\mathcal{L}[\hat \rho] =  -i[\hat H_S,\hat \rho]+\sum_{\alpha\beta} \gamma_{\alpha\beta}\left(\hat L_\beta\hat \rho\hat L^\dagger_\alpha-\frac{1}{2}\left\{\hat L_\alpha^\dagger \hat L_\beta,\hat \rho\right\}\right),
\end{equation}
where $\gamma_{\alpha\beta}$ defines a decay matrix,  $\hat L_{\alpha}$ are system jump operators associated with a model system-reservoir interaction, and $\hat H_S$ is the system Hamiltonian (undriven or driven HTC model). The matrix $\gamma_{\alpha\beta}$ can be diagonalized together with a redefinition of the jump operators in order to obtain a master equation in Lindblad (diagonal) form, which preserves positivity of $\hat\rho$ \cite{Breuer-book}. For jump operators $\hat L_\alpha$ and $\hat L_\alpha^\dagger$ that are eigenoperators of the system Hamiltonian $\hat H_S$ we have $[\hat H_S,\hat L_\alpha^\dagger \hat L_\beta]=0$, and we can write
\begin{equation}\label{eq:L0+L1}
\mathcal{L}[\hat \rho] =\left( \mathcal{L}_0+\mathcal{L}_1\right)[\hat \rho],
\end{equation}
$\mathcal{L}_0$ can be written as the non-Hermitian commutator 
 \begin{equation}\label{eq:L0}
 \mathcal{L}_0[\hat \rho] = -i\left(H_0\hat \rho-\hat\rho \hat H_0^\dagger\right),
 \end{equation}
 where
\begin{equation}\label{eq:Heff}
\hat H_0 = \hat H_S -\frac{i}{2}\sum_{\alpha\beta}\gamma_{\alpha\beta}\hat A_\alpha^\dagger \hat A_\beta,
\end{equation}
is an effective Hamiltonian with eigenstates $\ket{\bar\epsilon_j}$ that satisfy complex-eigenvalue equations $\hat H_0\ket{\bar\epsilon_j} = \bar\epsilon_j\ket{\bar\epsilon_j}$ and $\bra{\bar\epsilon_j}\hat H_0^\dagger = \bar\epsilon_j^*\bra{\bar\epsilon_j}$, with ${\rm Im}[\bar\epsilon_j]< 0$. This effective Hamiltonian is associated with the non-unitary evolution operator  
\begin{equation}\label{eq:U0}
\hat U_0(\tau) =  {\rm exp}[{-i\hat H_0 \tau}],
\end{equation}
such that the state norm decays with time as $\langle \bar\epsilon_j|\bar\epsilon_j\rangle_t ={\rm exp}[2{\rm Im[\bar\epsilon_j]}t]$.
\\

The  recycling term in Eq. (\ref{eq:L0+L1}) is given by 
\begin{equation}\label{eq:L1}
\mathcal{L}_1 [\hat \rho] = \sum_{\alpha\beta}\gamma_{\alpha\beta}\,\hat A_\beta\hat \rho\hat A_\alpha^\dagger, 
\end{equation}
and becomes the basis for our calculation of the emission and transmission spectra in organic cavities. When the system density matrix $\hat \rho(t)$ is such that the modified initial condition $\sigma_t(0)$ in Eq. (\ref{eq:sigma t}) satisfies, for every $t$, the equality
\begin{equation}\label{eq:sigma condition}
\mathcal{L}_1[\hat \sigma_t(0)]= 0,
\end{equation}
then the two-point correlation function in Eq. (\ref{eq:TPC sigma}) can be written as
\begin{eqnarray}\label{eq:TPC U0}
\langle\hat O_1(t+\tau)\hat O_2(t)\rangle &=& {\rm Tr}_S\left\{\hat O_1\,{\rm e}^{\mathcal{L}_0\tau}\hat \sigma_t(0)  \right\}\\
&=&{\rm Tr}_S\left\{\hat O_1 \hat U_0(\tau)\hat \sigma_t(0)\hat U_0^\dagger(\tau)\right\},\nonumber
\end{eqnarray}
Expanding the modified initial density matrix $\hat \sigma_t(0)$ in the eigenstates of $\hat H_0$ as
\begin{equation}\label{eq:sigma eigenexpansion}
\hat \sigma_t(0) = \sum_{ij}\sigma_{ij}(t)\ket{\bar\epsilon_i}\bra{\bar\epsilon_j},
\end{equation}
allows us to write Eq. (\ref{eq:TPC U0}) in the simple form
\begin{equation}\label{eq:TPC final}
\langle\hat O_1(t+\tau)\hat O_2(t)\rangle = \sum_{ij}\langle \epsilon_i|\hat O_1|\epsilon_j\rangle \,{\rm e}^{-i\omega_{ij}\tau+\kappa_{ij}\tau}\sigma_{ij}(t),
\end{equation}
where $\omega_{ij} = {\rm Re}[\bar\epsilon_i]-{\rm Re}[\bar\epsilon_j]$ and $\kappa_{ij} = {\rm Im}[\bar\epsilon_i]+{\rm Im}[\bar\epsilon_j]$. The corresponding spectrum of fluctuations gives
\begin{eqnarray}\label{eq:S12}
S_{12}(\omega,t)&=&\,{\rm Re}\int_0^\infty d\tau \langle \hat O_1(t+\tau)\hat O_2(t)\rangle {\rm e}^{i\omega \tau}\\
&=& \sum_{kij}\rho_{kj}(t)\langle \epsilon_k|\hat O_2^\dagger|\epsilon_i\rangle \langle \epsilon_i|\hat O_1|\epsilon_j\rangle\frac{\kappa_{ij}}{(\omega-\omega_{ij})^2+\kappa_{ij}^2},\nonumber
\end{eqnarray}
where in the last line we used
\begin{equation}
\sigma_{ij}(t) = \langle \epsilon_i|\hat O_2\hat \rho(t)|\epsilon_j\rangle = \sum_k \langle \epsilon_i|\hat O_2|\epsilon_k\rangle\rho_{kj}(t),
\end{equation}
with $\rho_{kj}(t) = \langle \epsilon_k|\hat \rho(t)|\epsilon_j\rangle$ being an element of the system density matrix in the eigenbasis of $\hat H_0$. In the steady state ($t\rightarrow \infty$), the fluctuations become stationary and the spectrum becomes independent of the absolute time $t$. It is reasonable to assume that in the steady state the system density matrix $\hat \rho$ becomes diagonal in the energy eigenbasis, i.e., $\hat \rho = \sum_j\rho_j \ket{\epsilon_j}\bra{\epsilon_j}$. If in addition to the mixed character of the system state, the fluctuation operators satisfy $\hat O_1 = \hat O_2^\dagger$, which is the relevant case for cavity emission and absorption, the spectrum of fluctuations in Eq. (\ref{eq:S12}) simplifies even further to read
\begin{eqnarray}\label{eq:steady spectrum}
S (\omega) = \sum_{ij}\rho_j |\langle \epsilon_i|\hat O_2|\epsilon_j\rangle|^2 \frac{\kappa_{ij}}{(\omega-\omega_{ij})^2+\kappa_{ij}^2},
\end{eqnarray}
which is the form used in the main text to model the spectra of organic microcavities, under the approximation that ${\rm Im}[\epsilon_i]\ll {\rm Im}[\epsilon_j]$ when $\ket{\epsilon_i}$ is in the ground state manifold and $\ket{\epsilon_j}$ is a polariton eigenstate of the Holstein-Tavis-Cummings model. For the Lindblad quantum master equation in Eq. (\ref{eq:LQME}), we thus have 
\begin{equation}\label{eq:Imj} 
\kappa_{ij}\approx \Gamma_j/2 =  \frac{1}{2}\sum_i\gamma_{ij}.
\end{equation}

\subsection{Validity of the NQJ Approximation}
\label{app:validity}

 For our purposes, we specialize the discussion to system states $\hat \rho(t)$ that involve {\it at most} one electronic excitation or photon in the cavity, but any possible number of purely vibrational excitations. We assume that at time $t$ the system can be described by a state of the form
\begin{equation}\label{eq:rho ansatz}
\hat \rho(t) = \hat\rho^{(0)}(t)+\hat\rho^{(1)}(t),
\end{equation}
where 
\begin{equation}\label{eq:rho zero}
\hat \rho^{(0)}(t) = \sum_{ii'}\rho_{ii'}(t)\ket{\epsilon_i}\bra{\epsilon_{i'}},
\end{equation}
is the reduced density matrix of the system projected onto the ground state manifold $\ket{\epsilon_i}$, and 
\begin{equation}\label{eq:rho one}
\hat \rho^{(1)}(t) = \sum_{jj'}\rho_{jj'}(t)\ket{\epsilon_j}\bra{\epsilon_{j'}},
\end{equation}
into the polariton manifold $\ket{\epsilon_j}$, following the notation convention in the main text for the HTC eigenstates. The ground manifold eigenstates states have the form
\begin{equation}\label{eq:alpha zero}
\ket{\epsilon_i} =\ket{g_1g_2\ldots g_N}\ket{0_c}\ket{\{v\}}_i,
\end{equation}
where $\ket{0_c}$ is the vacuum state of the cavity and $\ket{\{v\}}_i=\ket{v_1v_2\ldots v_N}_i$ describes the vibrational eigenstate of each molecule in the ensemble in the electronic ground state $\ket{g}$. On the other hand, HTC polariton eigenstates can be written as
\begin{equation}\label{eq:alpha one}
\ket{\epsilon_j} = b_{j} \ket{\psi_{j}^{e}}\ket{0_c}+c_{j}\ket{\phi_{j}^{g}}\ket{1_c},
\end{equation}
 where $\ket{\psi_{j}^{e}}$ is an electron-vibration state with any number of vibrational excitations but at most one molecule in its excited electronic state $\ket{e}$, and $\ket{\phi_{j}^{g}}$ is a state with any number of vibrational excitations but with all molecules in the ground electronic state $\ket{g}$.  For a cavity field near resonant with the bare molecular transition frequency we have $|c_j|^2\approx |b_j|^2 \approx1/2$.
 
We now consider the action of the electronic lowering operator  $\sigma_n^-\equiv \ket{g_n}\bra{e_n}$, where $n$ labels a molecule in the ensemble, and the cavity photon annihilation operator $\hat a$ on the one-quantum energy eigenstates in the form of Eq. (\ref{eq:alpha one}). For a dipole transition we have 
 \begin{equation} \label{eq:sm one}
 \hat\sigma_n^- \ket{\epsilon_j^{(1)}} =  b_{j}\ket{\gamma_{j}^g}\ket{0_c}=b_j\sum_kv_{jk}\ket{\epsilon_k^{(0)}},
 \end{equation} 
 where we defined $\ket{\gamma_{j}^g} \equiv \sigma_n^-\ket{\psi_{j}^e}$. For the cavity photon operator we have
 \begin{equation} \label{eq:a one}
\hat a \ket{\epsilon_j^{(1)}} =  c_{j}\ket{\phi_{j}^g}\ket{0_c} = c_j\sum_k u_{jk}\ket{\epsilon_k^{(0)}}.
 \end{equation}
Equations (\ref{eq:sm one}) and (\ref{eq:a one}) show that the identities 
\begin{eqnarray}\label{eq:zero evolution}
\hat a^2\ket{\epsilon_j^{(1)}} =(\hat \sigma_n^-)^2\ket{\epsilon_j^{(1)}} = 0
\end{eqnarray}
hold for all one-quantum eigenstates $\ket{\epsilon_j}$ and molecule index $n$. Physically, since the operators $\hat \sigma_n$ and $\hat a$ remove one quantum of electronic of cavity field energy, respectively, from the light-matter system, leaving the state in the manifold spanned by the zero-quantum eigenstates $\ket{\epsilon_i}$, no further high-frequency de-excitations are allowed. 

We illustrate the use of Eq. (\ref{eq:zero evolution}) with an example. Consider the leakage photoluminescence (LPL) spectrum.  In this case we are interested in evaluating the two-point correlation function $\langle \hat a^\dagger(t+\tau)\hat a(t) \rangle$, for a dissipative cavity dynamics involving a recycling term of the form \cite{Carmichael-book1}
\begin{equation}
\mathcal{L}_1[\hat \rho] = \kappa\,\hat a \,\hat \rho \,\hat a^\dagger,
\end{equation}
where $\kappa>0$ is the decay rate for an empty cavity.  Condition (\ref{eq:sigma condition}) for $\hat O_2 = \hat a$ thus reads 
\begin{equation}
\mathcal{L}_1[\hat a\hat \rho] \equiv \kappa\, \hat a^2 \hat \rho \hat a^\dagger = \kappa\, \hat a^2 (\hat \rho^{(0)}+ \hat \rho^{(1)})\hat a^\dagger = 0,
\end{equation}
where we have inserted our one-quantum ansatz for $\hat \rho$ from Eq. (\ref{eq:rho ansatz}), and used Eq. (\ref{eq:zero evolution}). The validity of the condition in Eq. (\ref{eq:sigma condition}) allows us to write the stationary LPL spectrum as
\begin{equation}\label{eq:LPL spectrum}
S_{\rm LPL}(\omega) = \sum_{ij}\rho_j\,|\langle\epsilon_i| \hat a| \epsilon_j\rangle |^2 \frac{\kappa_{ij}}{(\omega-\omega_{ij})^2+\kappa_{ij}^2},
\end{equation}
which is a special case of Eq. (\ref{eq:steady spectrum}) for $\hat O_2 = \hat a$. 

 Contributions to the state $\hat \rho$ from two-quantum states $\hat \rho^{(2)}$ gives $\mathcal{L}_1[\hat a\hat \rho]\neq 0$. In this case, the derived formula for $S_{\rm LPL}(\omega)$ becomes inaccurate. Weak driving of the cavity, either coherent or incoherent, such that $\langle \hat a^\dagger \hat a\rangle\ll 1 $ ensures that the one-quantum ansatz for $\hat\rho$ is appropriate to interpret the observable LPL spectrum. 

\section{Absorption of a Weak Driving Field}
\label{app:weak absorption}

We consider the driven Holstein-Tavis-Cummings (HTC) Hamiltonian $\hat H(t) = \hat{\mathcal{H}}+\hat V_p(t)$, where $\hat{\mathcal{H}}$ is the undriven HTC model in Eq. (\ref{eq:HTC}) and 
\begin{equation}\label{eq:Vp}
\hat V_{\rm p}(t) = \Omega_p\left(\hat a^\dagger\,{\rm e}^{-i\omega_p t}+\hat a\,{\rm e}^{i\omega_p t}\right),
\end{equation}
is the periodic driving term with frequency $\omega_p$ and a small amplitude $\Omega_p\ll \sqrt{N}\Omega\ll \omega_c$. The quantum master equation for the system density matrix now reads $\dot{\hat \rho} = \mathcal{L}[\hat\rho]+\mathcal{L}_p[\hat \rho]$, where $\mathcal{L}[\hat \rho]$ is the time-independent Liouville super-operator defined in Eq. (\ref{eq:L0+L1}) and 
\begin{equation}\label{eq:Lp}
\mathcal{L}_p[\hat \rho] = -i[\hat V_p(t),\hat \rho],
\end{equation}
is treated as a perturbation to the system dynamics. We expand the system state $\hat \rho$ as a perturbative expansion in the amplitude $\Omega_p$ of the form
\begin{equation}\label{eq:rho expansion}
\hat \rho \approx \hat \rho^{(0)}+\hat \rho^{(1)}+\hat \rho^{(2)},
\end{equation}
 where $\hat \rho^{(n)}$ is  $O(\Omega_p^n)$ in the driving amplitude. Note the different notational meaning from Eq. (\ref{eq:rho ansatz}). We are interested in the population of the $j$-th polariton eigenstate to second order in the driving amplitude, i.e., $\rho_j^{(2)}=\bra{\epsilon_j}\hat \rho^{(2)}(t)\ket{\epsilon_j}$, subject to the initial condition
 \begin{equation}\label{eq:initial condition}
 \hat \rho(0)  = \ket{G}\bra{G}, 
 \end{equation}
where $\ket{G}$ is the absolute ground state of the system. This choice is appropriate to describe an undriven cavity at room temperature ($k_bT\ll \omega_{\rm v}$). 
 
 The zeroth order equations of motion in the HTC eigenbasis are given by Eq. (\ref{eq:LQME}), i.e.,
\begin{equation}\label{eq:QME zero}
\frac{d}{dt}\hat \rho^{(0)} = \sum_{ij} \frac{\gamma_{ij}}{2}\left(2 \ket{\epsilon_i}\bra{\epsilon_j} \hat \rho^{(0)}  \ket{\epsilon_j}\bra{\epsilon_i} - \left\{\ket{\epsilon_j}\bra{\epsilon_j},\hat \rho^{(0)} \right\}  \right),
\end{equation}
Since the initial condition in Eq. (\ref{eq:initial condition}) is a zero-energy eigenstate of the jump operator $\ket{\epsilon_i}\bra{\epsilon_j}$, the zero-th order solution is $\rho_j^{(0)} = 0=\rho_{ij}^{(0)}$ and $\rho_{i} = \delta_{Gi}$, where $\rho_{ij}= \bra{\epsilon_i}\hat \rho\ket{\epsilon_j}$ is an optical coherence  and $\rho_i=\bra{\epsilon_i}\hat\rho\ket{\epsilon_i}$ a vibrational population in the ground state manifold. 

We then proceed to the first order equations of motion, which read
\begin{eqnarray}\label{eq:first order}
\dot \rho_i^{(1)} &=&\sum_{ij}\gamma_{ij}\rho_j^{(1)}\\
\dot \rho_j^{(1)} &=&-\sum_{ij}\gamma_{i}\rho_j^{(1)}\\
\dot \rho_{ij}^{(1)} &=&-\kappa_{ij}\rho_{ij}^{(1)}+i\delta_{iG}V_{ij}(t),
\end{eqnarray}
where $\kappa_{ij} \approx \sum_{i'}\gamma_{i' j}/2$, ignoring the decay of states $\ket{\epsilon_i}$ in the ground state manifold. The coherence equation carries a contribution from the perturbation in Eq. (\ref{eq:Lp}), with $V_{ij}(t) = \bra{\epsilon_i}\hat V_p(t)\ket{\epsilon_j}$. Given the initial condition in Eq. (\ref{eq:initial condition}), the equation for the first order coherence can be formally solved to give
\begin{equation}\label{eq:coherence one}
\rho_{ij}^{(1)}(t)  =i\,\delta_{iG} \int_0^t d\tau\, {\rm e}^{-\kappa_{ij}(t-\tau)}\, V_{ij}(\tau).
\end{equation}
 Proceeding to second order, we can write the polariton population equation as
 \begin{eqnarray}\label{eq:second order}
 \dot\rho_j^{(2)}(t)= -\sum_i\gamma_{ij}\rho_j^{(2)}(t)+i2\,{\rm Im}\{V_{jG}(t)\rho_{Gj}^{(1)}(t)\},
 \end{eqnarray}
which can be formally solved using Eq. (\ref{eq:coherence one}) to read
\begin{equation}\label{eq:rho two}
\rho_j^{(2)}(t) = {\rm e}^{-\Gamma_j}\rho_j^{(2)}(t)+\int_0^t dt' {\rm e}^{-\Gamma_j(t-t')} K(t'),
\end{equation}
where $\Gamma_j = \sum_{ij}\gamma_{ij}$ is the polariton decay rate and we defined the second-order kernel
\begin{equation}\label{eq:kernel}
K(s) = 2\,{\rm Re}\left\{\int_0^s d\tau\, {\rm e}^{-\kappa_{Gj}(s-\tau)}\, V_{jG}(s)V_{Gj}(\tau)\right\}.
\end{equation} 

In the steady-state ($t\rightarrow \infty$), the first term in Eq. (\ref{eq:rho two}) vanishes for all initial states $\hat\rho(0)$, so we ignore it. Inserting the definition in Eq. (\ref{eq:Vp}) and integrating the kernel $K(t')$ gives the polariton population
\begin{eqnarray}
\rho_j^{(2)}(t) &=& 2|\Omega_p|^2|\bra{G}\hat a\ket{\epsilon_j}|^2{\rm Re}\left\{\frac{1}{i\alpha_j}\left[\frac{1}{\Gamma_j}-\frac{{\rm e}^{-\Gamma_j t}}{\Gamma_j}\right.\right.\nonumber\\
&&\left.\left.-\left(\frac{{\rm e}^{(\Gamma_j-i\alpha_j) t}}{\Gamma_j-i\alpha_j}-\frac{1}{\Gamma_j-i\alpha_j}\right)\right]\right\} {\rm e}^{-\Gamma_j t},
\end{eqnarray}
where $\alpha_j \equiv (\omega_p-\omega_{Gj})-i\kappa_{Gj} $. In the steady-state limit $t\rightarrow \infty$, only the first term in square brackets survives, giving the desired stationary polariton population
\begin{equation}\label{eq:rho steady}
\rho_j^{(2)}(\infty) = \frac{2|\Omega_p|^2}{\Gamma_j}|\bra{G}\hat a\ket{\epsilon_j}|^2\frac{\kappa_{Gj}}{(\omega_p-\omega_{Gj})^2+\kappa_{Gj}^2}.
\end{equation}

The stationary polariton population $\rho_j$ enters in the expression for the bound fluorescence spectra from Eq. (\ref{eq:steady spectrum}), with $\hat O_2 = \hat \mu^{(+)}$, which reads
\begin{equation}\label{eq:S mu}
S_{\hat \mu}(\omega) = \sum_{ij}\rho_j |\langle \epsilon_i|\hat \mu^{(+)}|\epsilon_j\rangle|^2 \frac{\kappa_{ij}}{(\omega-\omega_{ij})^2+\kappa_{ij}^2}.
\end{equation}
For $\rho_j$ given by Eq. (\ref{eq:rho steady}), we integrate $S_{\hat \mu}$ over all emission frequencies $\omega$ to obtain the absorption spectra $A(\omega_p)$ from Eq. (\ref{eq:absorption def}) to read
\begin{equation} \label{eq:A omegap}
\frac{A(\omega_p)}{|\Omega_p|^2}  = \sum_{ij}\left(\frac{\pi}{\Gamma_j}\right)\frac{|\langle G|\hat a|\epsilon_j\rangle|^2\,\kappa_{Gj}}{(\omega_p-\omega_{Gj})^2+\kappa_{Gj}^2}|\langle\epsilon_i|\hat \mu^{(+)}| \epsilon_j\rangle|^2,
 \end{equation}
where we have used a Lorentzian normalization relation. For decoherence due to radiative decay only, we have $\kappa_{Gj}\approx \Gamma_j/2$. However, we can allow $\kappa_{Gj}$ to be more general and account also for coherence decay due to vibrational relaxation.

\end{document}